\documentclass[a4paper,11pt]{article}
\usepackage{jinstpub} 
\usepackage{lineno}
\usepackage{subcaption}
\usepackage[LGRgreek]{mathastext}
\title{\boldmath Resistive Plate Chamber Detector Construction and Certification: State-of-the-Art Facilities at the Max Planck Institute for Physics, in Partnership with Industrial Partners}

\author[a,b]{Davide Costa,}
\author[a,1]{Francesco Fallavollita,\note{Corresponding author.}}
\author[a,b]{Hubert Kroha,}
\author[a,b]{Oliver Kortner,}
\author[a]{Pavel Maly,}
\author[a]{Giorgia Proto,}
\author[a]{Daniel Soyk,}
\author[a]{Elena Voevodina,}
\author[a]{Jorg Zimmermann}

\affiliation[a]{Max-Planck-Institut für Physik, \\ Werner Heisenberg Institute, \\ Boltzmannstr. 8 - 85748 Garching, \\ Germany}
\affiliation[b]{Physik-Department, \\ Technische Universität München, \\ James-Franck-Str. 1 - 85748 Garching, \\ Germany}

\emailAdd{fallavol@mpp.mpg.de}

\abstract{
Resistive Plate Chambers (RPCs) featuring 1 mm gas volumes combined with high-pressure phenolic laminate (HPL) electrodes provide excellent timing resolution down to a few hundred picoseconds, along with spatial resolution on the order of a few millimeters. Thanks to their relatively low production cost and robust performance in high-background environments, RPCs have become essential components for instrumenting large detection areas in high-energy physics experiments.
The growing demand for these advanced RPC detectors, particularly for the High-Luminosity upgrade of the Large Hadron Collider (HL-LHC), necessitates the establishment of new production facilities capable of delivering high-quality detectors at an industrial scale. To address this requirement, a dedicated RPC assembly and certification facility has been developed at the Max Planck Institute for Physics in Munich, leveraging strategic collaborations with industrial partners MIRION\textsuperscript{\tiny®} and PTS\textsuperscript{\tiny®}. This partnership facilitated the transfer of advanced, research-level assembly methodologies into robust, scalable industrial processes. Central to this development are innovations such as automated precision assembly systems and comprehensive quality assurance protocols, integrated directly into the production workflow to ensure adherence to stringent performance standards required by experiments like ATLAS. Through a structured, phased prototyping and certification approach, initial tests on small-scale ($40 \times 50 \, cm^2$) prototypes validated the scalability and applicability of optimized production procedures to large-scale ($1.0 \times 2.0 \, m^2$) RPC detectors. Currently, the project has entered its final certification phase, involving extensive performance and longevity testing, including a year-long irradiation campaign at CERN's Gamma Irradiation Facility (GIF++). This article details the development and successful industrial implementation of novel assembly techniques, highlighting the enhanced capabilities and reliability of RPC detectors prepared through this industrial-academic collaboration, ensuring readiness for upcoming challenges in high-energy physics detector instrumentation.}

\keywords{Only keywords from JINST's keywords list please}

\arxivnumber{1234.56789} 

\begin{document}
\maketitle
\flushbottom

\section{New facilities for scalable production of RPC detectors}
\label{sec:TestBeamStudies}
In response to the growing demand for high-rate Resistive Plate Chambers (RPCs) in high-energy physics experiments, new large-scale production facilities and methods have been developed. These advanced RPCs feature a 1‐mm‐thick gas volume with high-pressure phenolic laminate (HPL) electrodes, providing enhanced background rate capability and longevity, essential for experiments like those at the High-Luminosity LHC (HL-LHC) and beyond \cite{2106380}. To ensure high-quality, efficient production, a dedicated assembly and certification facility has been established at the Max Planck Institute for Physics in Munich, supported by strategic collaborations with two German industrial manufacturers, PTS\textsuperscript{\tiny \textregistered} and MIRION\textsuperscript{\tiny \textregistered} companies. This partnership enabled the seamless transfer of advanced assembly techniques from research to industry, incorporating innovations such as automated precision assembly lines and rigorous quality control protocols embedded within the production workflow to ensure both scalability and compliance with stringent ATLAS performance standards. The prototyping and certification process followed a structured, phased approach to systematically prepare for large-scale production of the gas volumes, ensuring compliance with stringent industrial standards \cite{Kortner:2023tob}. In the initial phase, the developments have been based on the production procedures initially employed by the ATLAS and CMS experiments for their RPC detector technology. These procedures have been subsequently refined to facilitate production at an industrial scale. The optimized procedure has been implemented on $40 \times 50 \, cm^2$ small-scale RPC detector prototypes, demonstrating its scalability and applicability to full-scale RPCs with final dimensions of $1.0 \times 2.0 \, m^2$. These small-scale detector prototypes played a crucial role in assessing both performance and manufacturing quality, offering valuable insights into the production and certification process. The collaboration is now advancing toward the final certification phase, which encompasses comprehensive testing of both small- and large-scale RPC detector prototypes, with a strong focus on longevity studies, including a year-long irradiation test at CERN’s Gamma Irradiation Facility (GIF++). This phase is pivotal in confirming that the selected manufacturers meet the stringent qualifications essential for large-scale production. This article will detail the novel assembly methods, the successful technology transfer to industrial partners, and the capacity of the new facilities to deliver high-quality RPCs on an industrial scale, ensuring readiness for high-energy physics applications beyond the HL-LHC upgrade.

\section{Test-beam studies of the small-scale RPC detector prototypes}
\label{sec:TestBeamStudies}
As part of the qualification process for large-scale production of RPC gas volumes for the ATLAS Muon Spectrometer upgrade, small-scale 1-mm gap RPC detector prototypes ($40 \times 50 \, cm^2$) have been successfully manufactured by the PTS\textsuperscript{\tiny \textregistered} and MIRION\textsuperscript{\tiny \textregistered} companies. In dedicated test-beam campaigns at CERN GIF++ facility, these prototypes have been subjected to comprehensive performance evaluations under conditions simulating the operational environment, with exposure to a $\sim$11 TBq $^{137}Cs$ source, emitting 662 keV gamma rays, in combination with a muon beam having a mean energy of approximately 100 GeV from the secondary SPS beam line H4 \cite{Pfeiffer:2016hnl}.  In the following, the main RPC detector prototype performances is reported in terms of muon detection efficiency, time resolution and radiation-induced current. For the standard ATLAS-RPC gas mixture (94.7\% $C_2H_2F_4$ : 5\% $i-C_4H_{10}$ : 0.3\% $SF_6$), the results are compared at different levels of irradiation, expressed by means of the gamma cluster rate as measured at the detector working point of 5.8 kV. The dependencies of relevant variables are presented as functions of the effective high voltage, $V_{eff}$, which is defined by the expression: 
\begin{equation}
V_{eff} = V_{app} \times \frac{P_0}{P} \times \frac{T}{T_0}
\label{eq:PTCorrectioGIF++}
\end{equation}

\noindent where $V_{app}$ is the applied high voltage, P represents the atmospheric pressure, and T the temperature at the time of data-taking. The constants $P_0$ = 990 mbar and $T_0$ = 293 K correspond to the average pressure and temperature conditions observed at the GIF++ facility. Further details and technical aspects regarding the experimental setup and analysis of test beam data obtained with small-scale RPC detector prototypes, produced by German manufacturers, are provided in \cite{Turkovic:2023yzk}. The following results validate the high manufacturing standards achieved by the German PTS\textsuperscript{\tiny \textregistered} and MIRION\textsuperscript{\tiny \textregistered} companies, underscoring their technical readiness to advance to large-scale production that fulfills the stringent performance and reliability requirements of the ATLAS Muon Spectrometer upgrade. 

\subsection{Muon Detection Efficiency}
\label{subsec:MuonDetectionEfficiency}
The muon detection efficiency is evaluated as the number of events where at least one cluster has been detected inside the muon window divided by the total number of triggers. The efficiency as a function of $V_{eff}$ is shown in Figure \ref{fig:MuonDetectionEfficiencyGIF++}  for the standard ATLAS-RPC gas mixtures at different gamma cluster rates evaluated at the detector working point. The efficiency dependency on $V_{eff}$ has been interpolated by means of the sigmoid functions: 
\begin{equation}
\epsilon = \frac{\epsilon_{max}}{1+e^{-\lambda(V_{eff} \, - \, V_{50\%})}}
\label{eq:SigmoidFunctionGIF++}
\end{equation}

\noindent where the parameter $\epsilon_{max}$ represents the maximum plateau efficiency, $V_{50\%}$ is the $V_{eff}$ at 50\% of the maximum efficiency and $\lambda$ is proportional to the slope of the efficiency curve at $V_{50\%}$. The RPC detector prototypes, operated with the standard gas mixture at the working point, demonstrated an average muon detection efficiency of approximately 97\% under source-off conditions. When exposed to medium particle rates, reaching up to $\sim1.3 \, kHz/cm^2$, this efficiency showed a slight reduction of about 3\%. At higher rates, up to $\sim2.4 \, kHz/cm^2$, a more pronounced degradation was observed, with efficiency decreasing by around 8\% and accompanied by a progressive shift in the working point toward higher values. Despite the rate-dependent effects observed, the RPC detector prototypes demonstrated considerable resilience, achieving a muon detection efficiency greater than 96.5\% at the nominal operating voltage at typical background rates in the HL-LHC runs ($\sim200-300 \, Hz/cm^2$).

\subsection{Time Resolution}
\label{subsec:TimeResolution}
The absolute time resolution of the RPC detector prototypes has been evaluated using the Time-of-Flight (TOF) method by measuring the difference in signal arrival times between pairs of detectors selected from a set of three. All possible detector combinations have been analyzed, consistently yielding comparable results, thereby validating the reliability and reproducibility of the measurements. The absolute time resolution has been determined by dividing the estimated time difference resolution by $\sqrt{2}$, under the assumption that both the detectors and electronic components of both detectors are identical and their contributions are uncorrelated. The width of time difference distributions $\langle \sigma_{\Delta t} \rangle$ is extracted using a Gaussian function and the absolute time resolution $\langle \sigma_t \rangle$ is given by:
\begin{equation}
\langle \sigma_t \rangle = \frac{\langle \sigma_{\Delta t} \rangle}{\sqrt{2}}
\label{eq:TimeResolutionGIF++}
\end{equation}

\noindent A preliminary result for the time difference distribution between signals generated by the same muons in two strips in two parallel gap RPC detector prototypes in coincidence is shown in Figure \ref{fig:TimeResolutionGIF++}. The standard deviation of the signal time difference distribution has been measured to be $573 \pm 13$ ps, corresponding to an absolute time resolution of $405 \pm 9$ ps for 100 GeV muon beam, well within the ATLAS requirement of 1 ns, demonstrating excellent timing performance of the prototypes.

\subsection{Gas Gap Current}
\label{subsec:Gas_Gap_Current}
\begin{figure}[h!]
    \centering
    \begin{subfigure}[b]{0.45\textwidth}
        \centering
        \includegraphics[width=\textwidth]{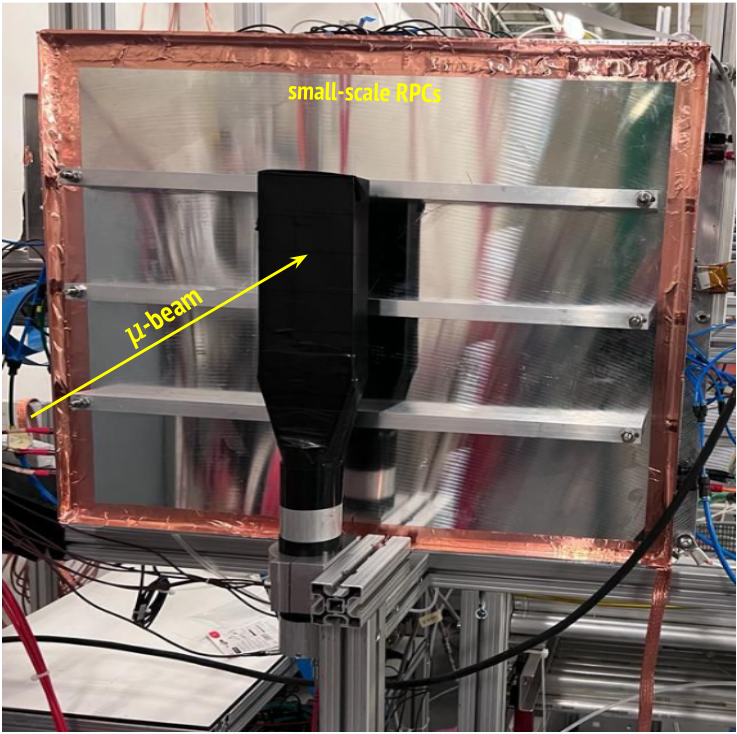}
        \caption{ }
        \label{fig:1a}
    \end{subfigure}
    \hfill
    \begin{subfigure}[b]{0.45\textwidth}
        \centering
        \includegraphics[width=\textwidth]{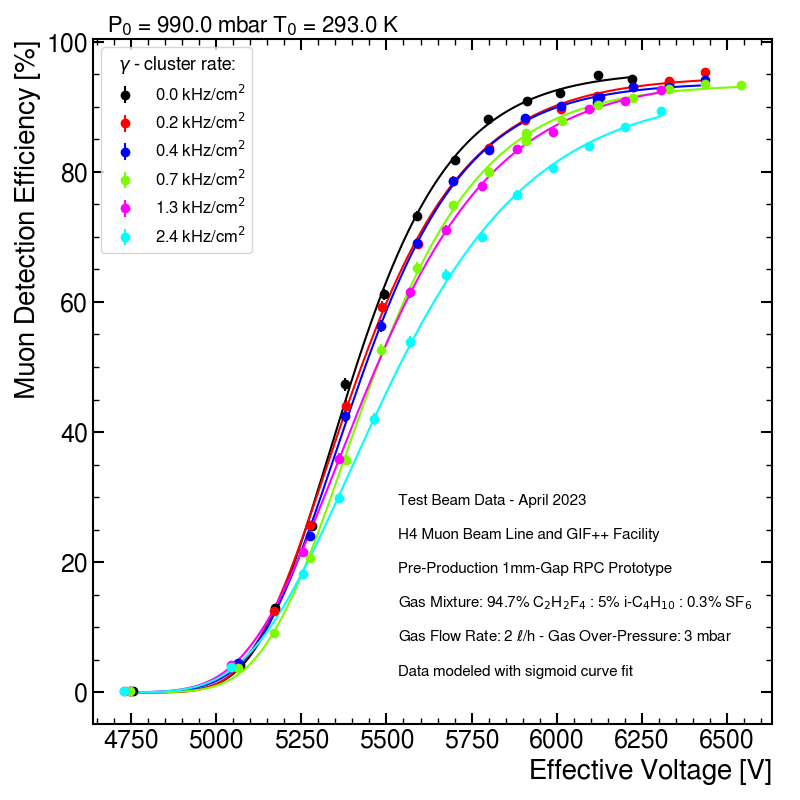}
        \caption{ }
        \label{fig:MuonDetectionEfficiencyGIF++}
    \end{subfigure}

    \vspace{0.5cm} 

    \begin{subfigure}[b]{0.45\textwidth}
        \centering
        \includegraphics[width=\textwidth]{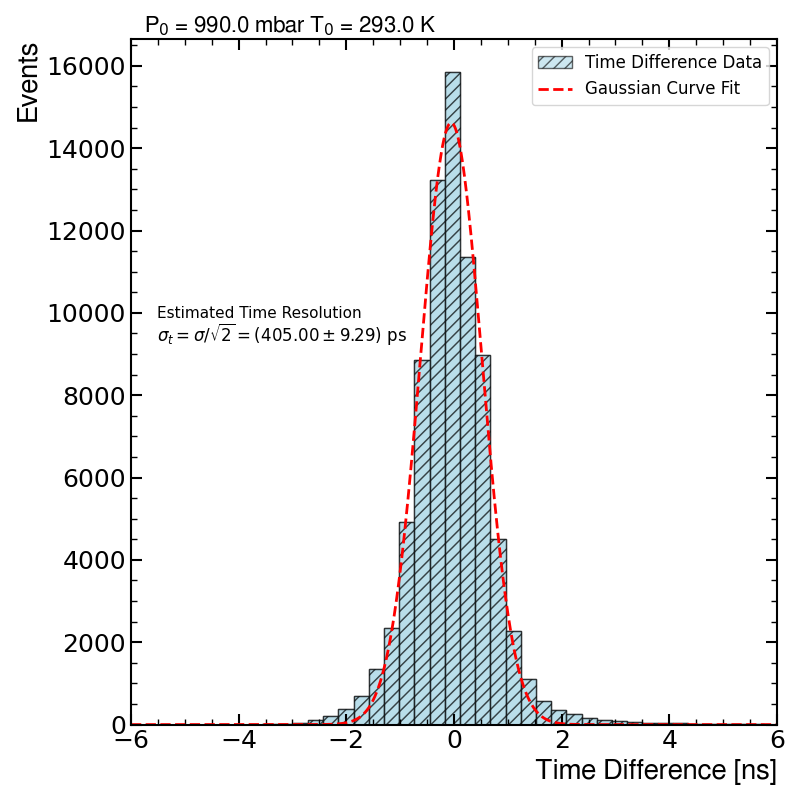}
        \caption{ }
        \label{fig:TimeResolutionGIF++}
    \end{subfigure}
    \hfill
    \begin{subfigure}[b]{0.45\textwidth}
        \centering
        \includegraphics[width=\textwidth]{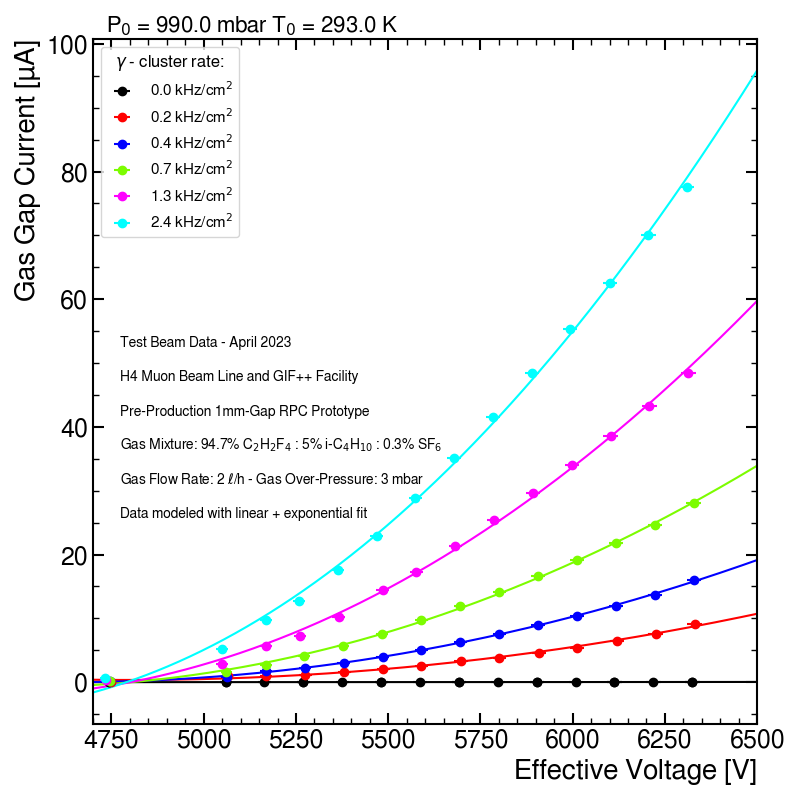}
        \caption{ }
        \label{fig:GasGapCurrentGIF++}
    \end{subfigure}
    
    \caption{(a) Photograph of the experimental setup installed at the CERN GIF++ facility. (b) Muon detection efficiency as a function of effective high voltage for the standard gas mixture. The efficiency data points are interpolated by a sigmoid function. (c) Time difference distribution and time resolution results of the beam test. The blue histogram represents the time difference distribution measured by the two RPC prototypes, and the red line is the result of the Gaussian fit. 
    (d) Current as a function of the effective high voltage. Results are shown at different gamma cluster rates evaluated at the detector working point}
    \label{TestBeamResultGIF++}
\end{figure}

A critical operational limitation of RPC detectors under irradiation is managing the current flowing through the gas volume. Excessive current could lead to potential damage, emphasizing the need for close monitoring to ensure the long-term reliability and safety of RPC detectors in high-radiation environments. To address this, current flow within the gas volume has been systematically monitored during both in-spill and out-of-spill data acquisition phases. Out-of-spill data has been specifically utilized to assess the average current, allowing for an evaluation of its mean value and associated statistical uncertainty. The current has been determined by measuring the voltage drop across a 100 k$\Omega$ resistor connecting one of the two HPL plates to ground. The radiation-induced currents as a function of the effective high voltage applied across the gas volume for standard ATLAS-RPC gas mixtures under different irradiation conditions are presented in Figure \ref{fig:GasGapCurrentGIF++}. 

\section{Assembly methodology for a full-scale RPC gas volume}
\label{sec:RPCAssemblyProcedure}
In this section, the production process for full-scale 1 mm gap RPCs are detailed, with particular emphasis on modifications implemented to streamline and enhance manufacturability in an industrial environment.


\subsection{Manufacturing Process of High-Pressure Laminate Electrodes}
\label{subsec:HPLElectrodes}
\begin{figure}[ht!]
    \centering
    \begin{subfigure}{0.48\textwidth}
        \includegraphics[width=\linewidth]{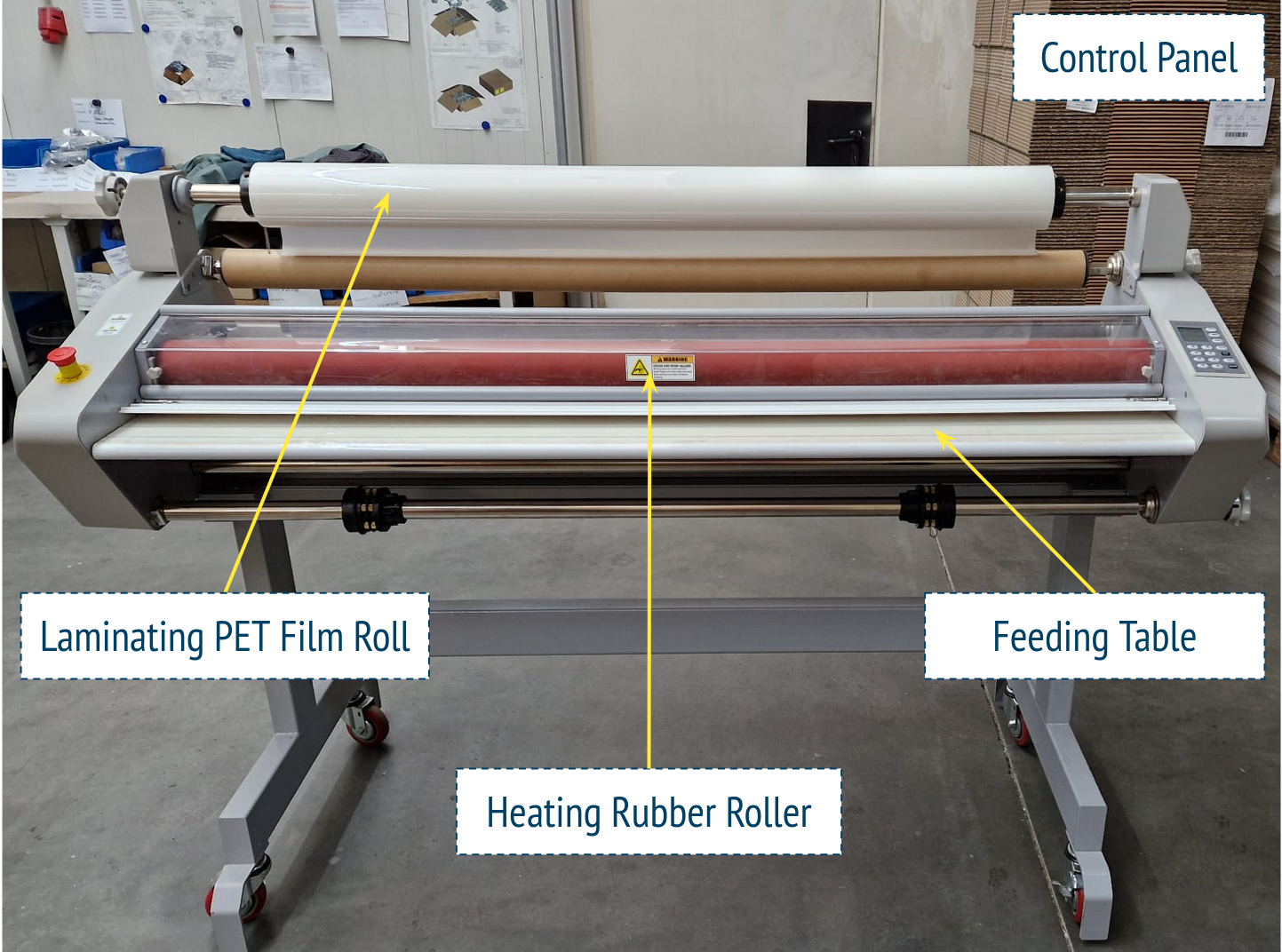}
        \caption{ }
        \label{fig:LaminatingPressMachine}
    \end{subfigure}
    \hfill
    \begin{subfigure}{0.48\textwidth}
        \includegraphics[width=\linewidth]{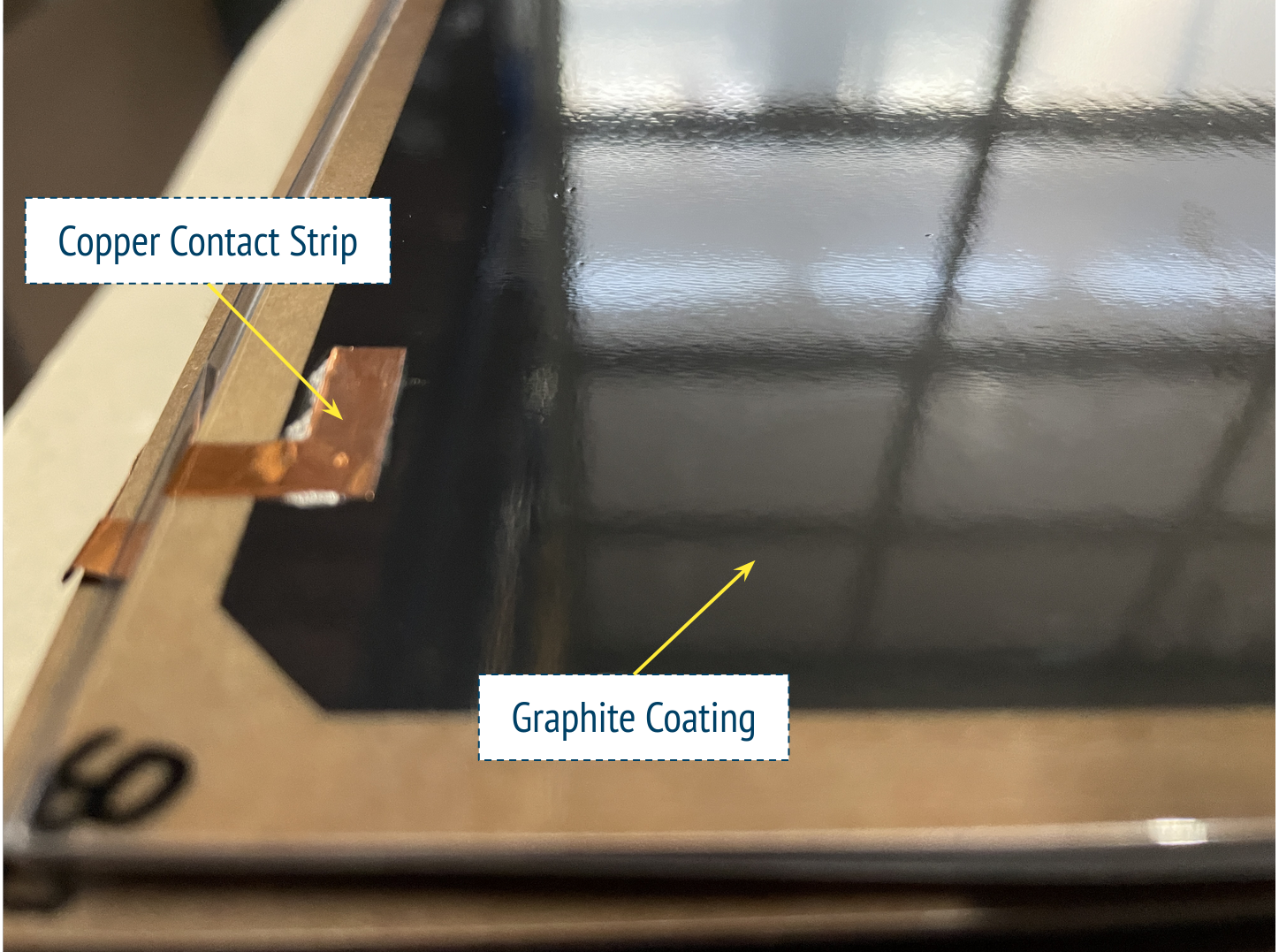}
        \caption{ }
        \label{fig:HPLElectrode}
    \end{subfigure}
    \caption{(a) Lamination press machine for the application of PET foil to graphite-coated HPL electrodes for electrical insulation. (b) HPL electrode following the completion of the PET lamination process.}
    \label{fig:GasVolumeAssemblyStep1}
\end{figure}

The RPC gas volume comprises two High Pressure Laminate (HPL) plates, each manufactured from multiple layers of kraft paper impregnated with phenolic resin. These plates are sourced from the Italian manufacturer Teknemica\textsuperscript{\tiny \textregistered} with strict quality specifications: the nominal thickness is specified as 1.40 mm, with a tolerance of +0.10 mm and -0.07 mm, while the tolerance on linear dimensions is $\pm 0.5$ mm. Additionally, to maintain uniformity and structural stability, the difference between the two diagonals is required to be less than 1 mm. The HPL plates feature a bulk resistivity range of $1.5 \times 10^{10} \, \Omega \, cm$ to $6 \times 10^{10} \, \Omega \, cm$, measured at 20$^{\circ}C$ and 50\% relative humidity, providing the necessary electrical characteristics for efficient detector operation and performance \cite{2106380}, \cite{Aielli:2016faq}. A precise graphite coating is applied to the external surfaces of the HPL plates to ensure consistent high-voltage distribution across the electrodes. This graphite layer, critical for the detector performance, has strict quality specifications: its surface resistivity is required to be 320 $k\Omega/\square$, with a tolerance of ±30\%, to minimize the voltage drop produced by the radiation-induced current, reduce the lateral dispersion of induced signals on the pick-up strips (thereby minimizing the muon cluster size), and maintain high electrode transparency to fast signals \cite{2106380}, \cite{Aielli:2016faq}. 
The application of this graphite varnish is achieved through a silk-screen printing process manufactured by the German company Siebdruck Esslinger\textsuperscript{\tiny \textregistered}, which has proven to be the most reliable method for meeting the specified surface resistivity across all electrodes. A copper contact strip is attached to the graphite-coated surface of the HPL electrodes with conductive silver adhesive to ensure optimal electrical connectivity. For effective electrical insulation in the HPL electrodes, a polyethylene terephthalate (PET) foil is laminated onto the graphite-coated surfaces of the electrodes. This insulation step involves bonding a 190 $\mu m$ thick PET foil, coated with 80 $\mu m$ thick low-melt ethylene vinyl acetate (EVA), to each electrode using a laminating pressing machine set to 105$^{\circ}C$ with lamination speed of $\sim3 \, m/min$. Figure \ref{fig:LaminatingPressMachine} and \ref{fig:HPLElectrode} show the laminating press machine and the HPL electrode upon the completion of the lamination process, respectively.

\subsection{Manufacturing Process of Gas Volume}
\label{subsec:GasGapAssembly}
\begin{figure}[ht!]
    \centering
    \begin{subfigure}{0.48\textwidth}
        \includegraphics[width=\linewidth]{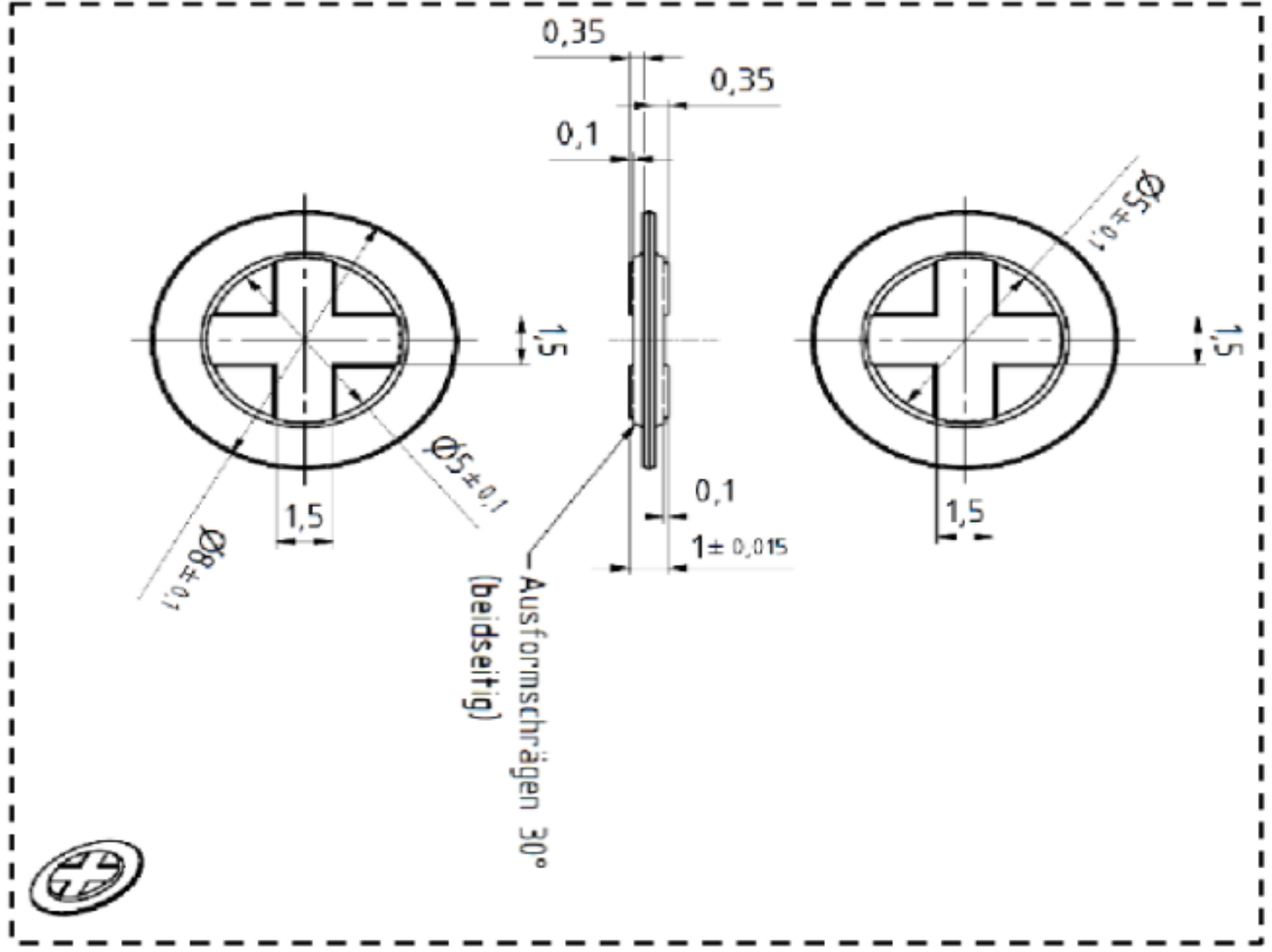}
        \caption{ }
        \label{fig:SpacerDrawings}
    \end{subfigure}
    \hfill
    \begin{subfigure}{0.48\textwidth}
        \includegraphics[width=\linewidth]{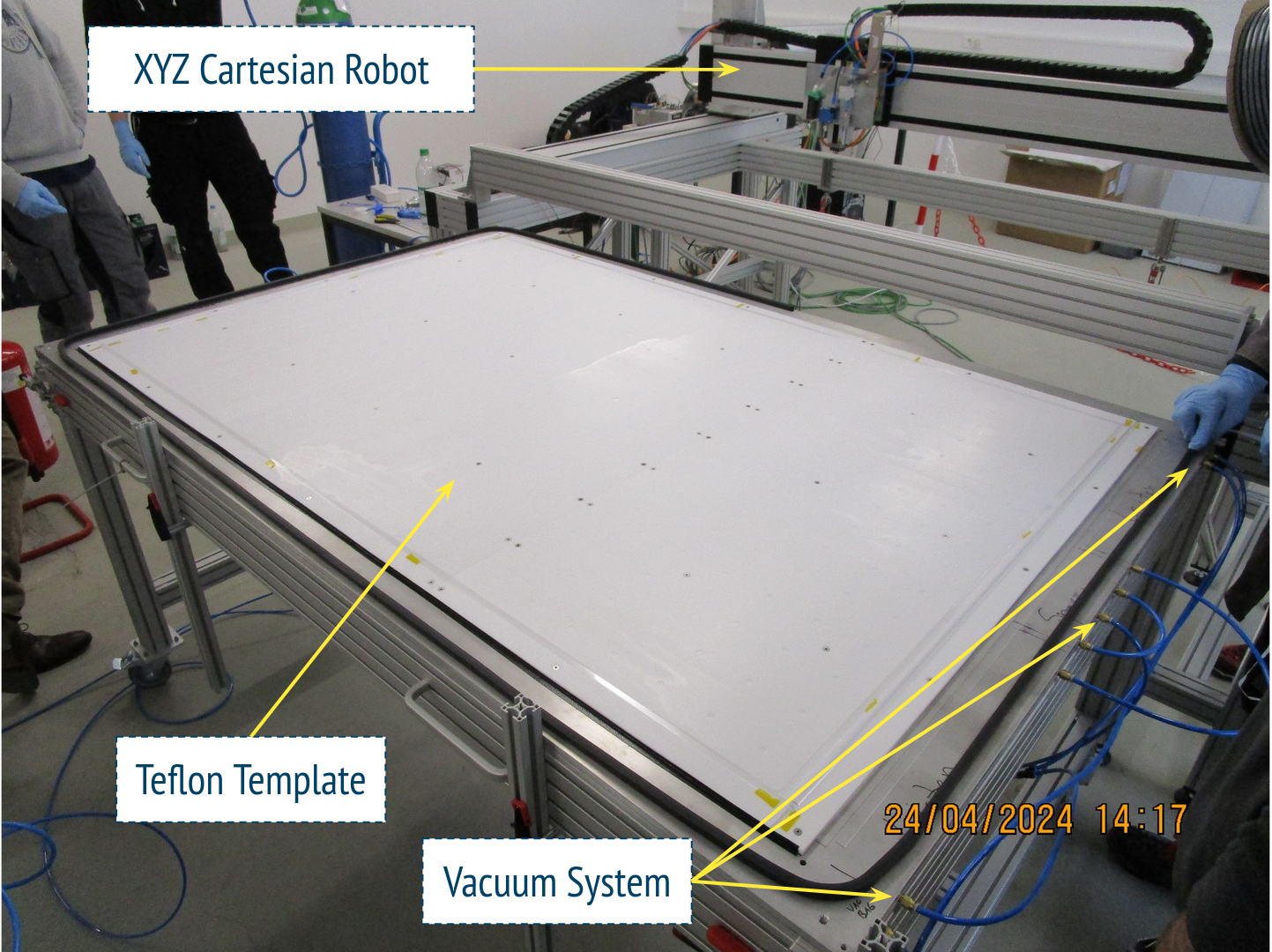}
        \caption{ }
        \label{fig:TeflonTemplate}
    \end{subfigure}
    \caption{(a) Technical drawing of the polycarbonate spacer, featuring stringent mechanical tolerances on thickness to ensure precise fitting. The crosswise dimple design establishes a controlled and reproducible adhesive gap, facilitating accurate and consistent assembly. (b) Teflon template used for precise positioning of poly-carbonate spacers and lateral profiles prior to gluing them onto the HPL resistive electrode.}
    \label{fig:GasVolumeAssemblyStep2}
\end{figure}

The two HPL electrodes are separated by mechanically precise poly-carbonate spacers, ensuring an accurately defined gas volume width. The mechanical precision required for the spacer thickness, with a tolerance of $\pm 15 \, \mu m$, is achieved through injection molding. To prevent gas volume variations exceeding $15 \, \mu m$, which could result from non-uniform glue thickness securing the spacers onto the HPL plates, and to ensure a consistently defined glue layer, a cross-shaped dimple was introduced into the spacer design, as shown in Figure \ref{fig:SpacerDrawings}. Additionally, gas tightness is maintained by a poly-carbonate lateral profile bonded along the perimeter, which also interfaces with the gas distribution system. The assembly of the gas volume demands precise timing, particularly when applying epoxy glue to bond poly-carbonate spacers and lateral profiles to the resistive electrodes. Given the limited working time of approximately 50 minutes, this phase introduces additional complexity, requiring careful coordination to achieve accurate alignment and secure bonding within the time constraints. To decouple the precise positioning of spacers, approximately 400 components per full-size gas volume, from the gluing process, a dedicated template has been developed, as shown in Figure \ref{fig:TeflonTemplate} This template comprises a Teflon plate with meticulously arranged slots and grooves, designed to accommodate both the spacers and lateral profiles. The Teflon plate is mounted on an aluminum frame equipped with a vacuum-based suction system, ensuring the secure placement of spacers and profiles during glue application and subsequent electrode alignment and positioning. Teflon has been chosen as the template material due to its non-stick properties, which prevent adhesive adherence and facilitate easy removal after assembly. Prior to assembly, the polycarbonate spacers, lateral profiles, and HPL plates are rigorously cleaned with isopropanol to ensure that the surfaces are free of contaminants and meet stringent cleanliness specifications.

\subsubsection{Gluing of the electrodes onto the spacers and lateral profiles}
\label{subsubsec:GasGapAssembly}
\begin{figure}[ht!]
    \centering
    \begin{subfigure}{0.48\textwidth}
        \includegraphics[width=\linewidth]{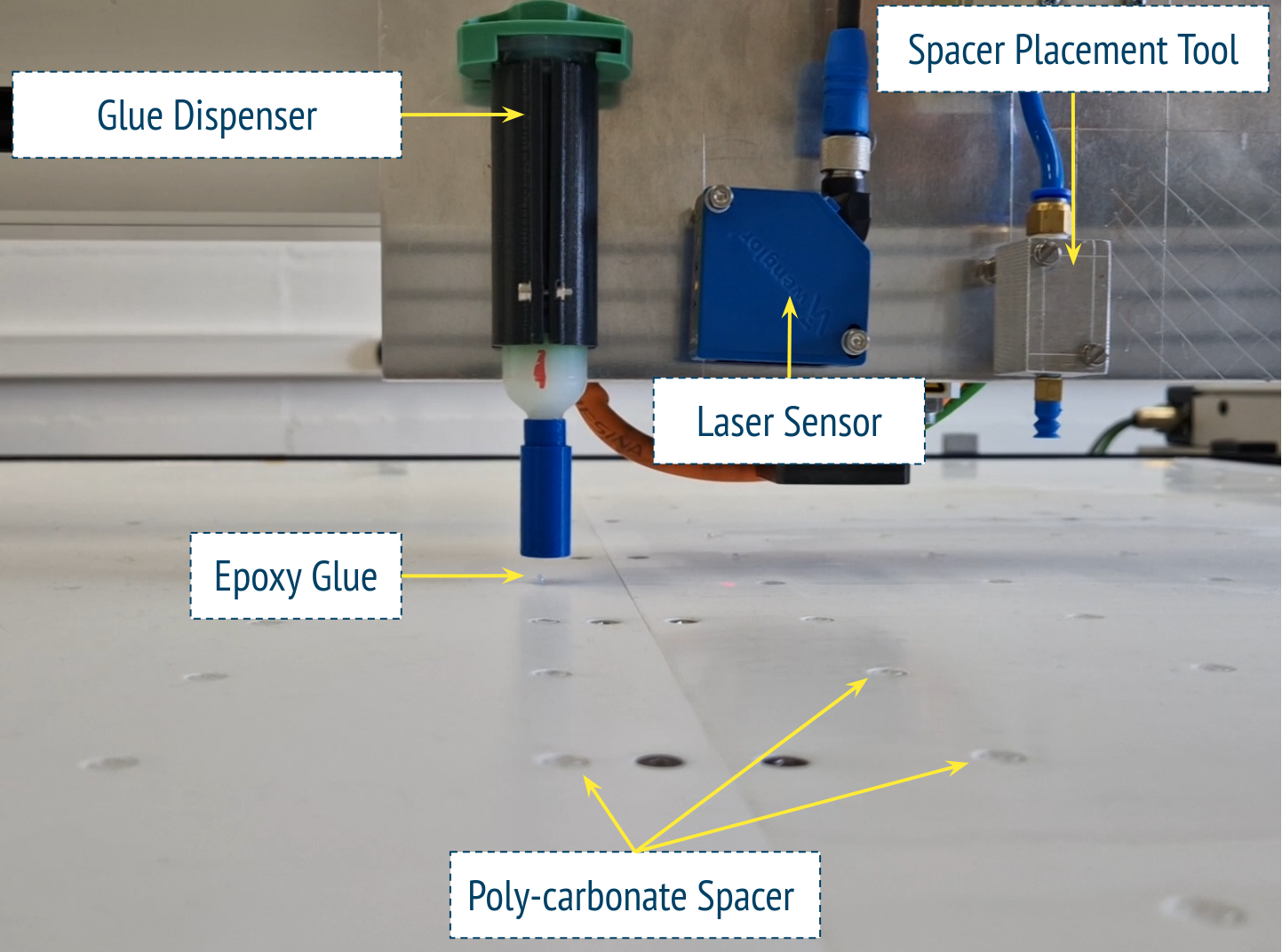}
        \caption{ }
        \label{fig:GlueDispensingSpacer}
    \end{subfigure}
    \hfill
    \begin{subfigure}{0.48\textwidth}
        \includegraphics[width=\linewidth]{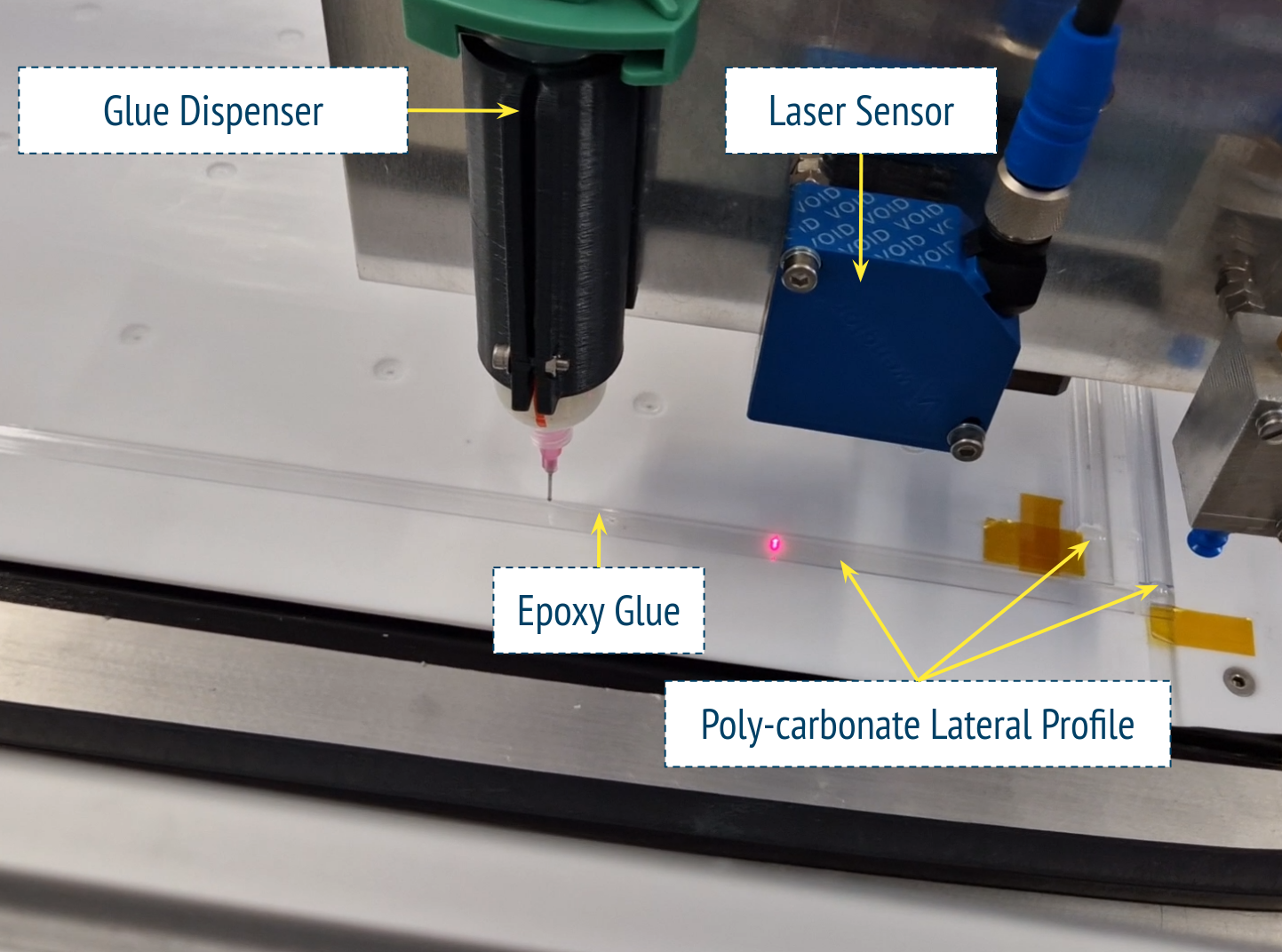}
        \caption{ }
        \label{fig:GlueDispensingLateralProfile}
    \end{subfigure}
    \caption{(a) and (b) Freeze frames capturing the glue dispensing onto the spacers (left)
and along the lateral profiles (right) of the gas volume.}
    \label{fig:GasVolumeAssemblyStep3}
\end{figure}

Following the precise placement of spacers and lateral profiles on the Teflon template, an automated glue dispenser, integrated with an XYZ Cartesian positioning systems, applies epoxy glue with high accuracy. The glue dispensing parameters, including dispensing pressure and time, and needle size, have been optimized to ensure the precise delivery of approximately $5 \, \mu l$ of glue for each spacer and to compensate for the increase in glue viscosity over time. This configuration ensures uniform and controlled glue distribution across all components, significantly improving assembly precision and structural integrity. Figures \ref{fig:GlueDispensingSpacer} and \ref{fig:GlueDispensingLateralProfile} show the freeze frames capturing the glue dispensing process during the assembly phase. 

\begin{figure}[h!]
    \centering
    \begin{minipage}{0.48\textwidth}
        \centering
        \includegraphics[width=\textwidth]{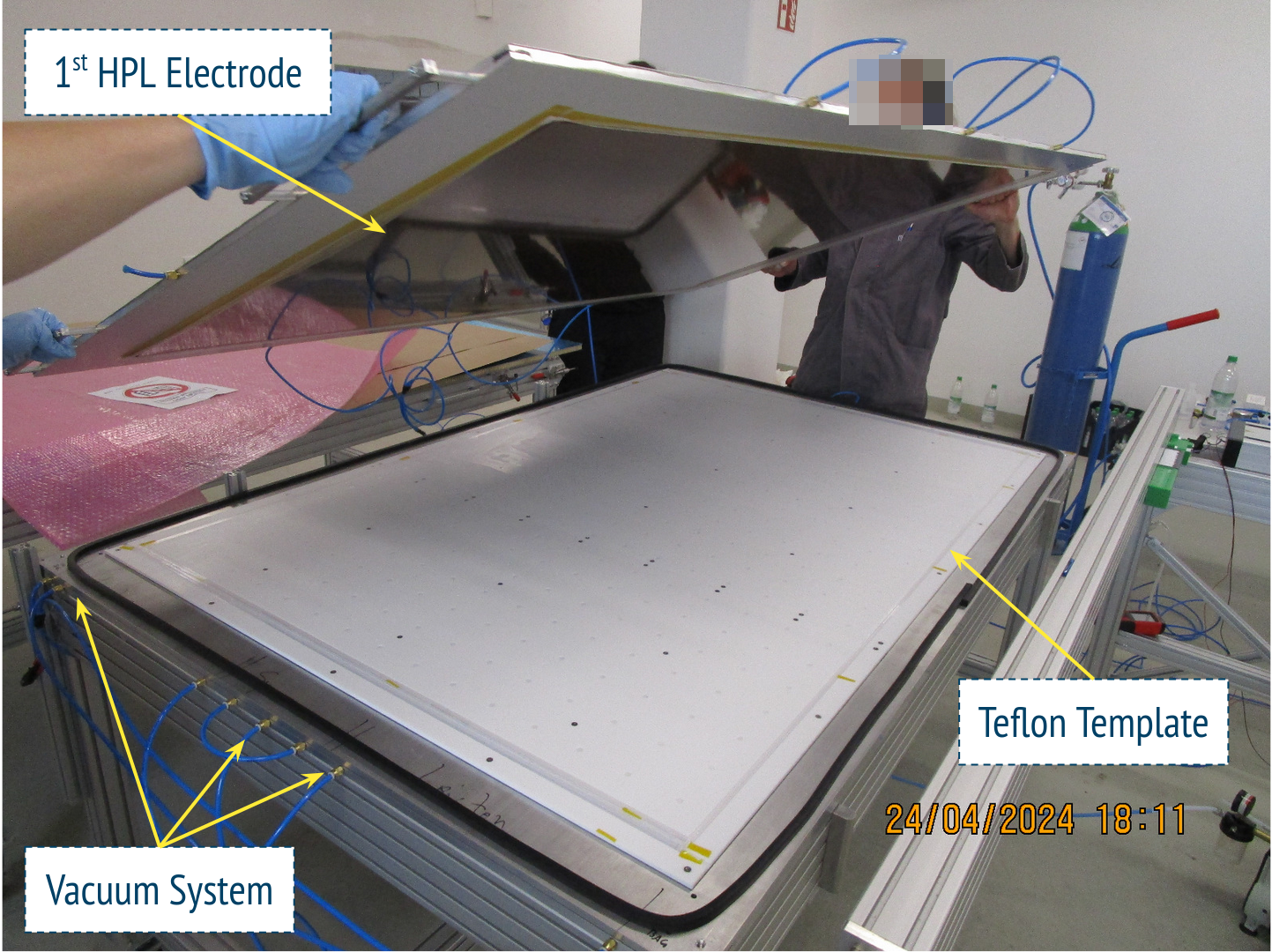} 
        \subcaption{ }
        \label{fig:1stHPLElectrodeInstallation}
    \end{minipage}
    \hfill
    \begin{minipage}{0.48\textwidth}
        \centering
        \includegraphics[width=\textwidth]{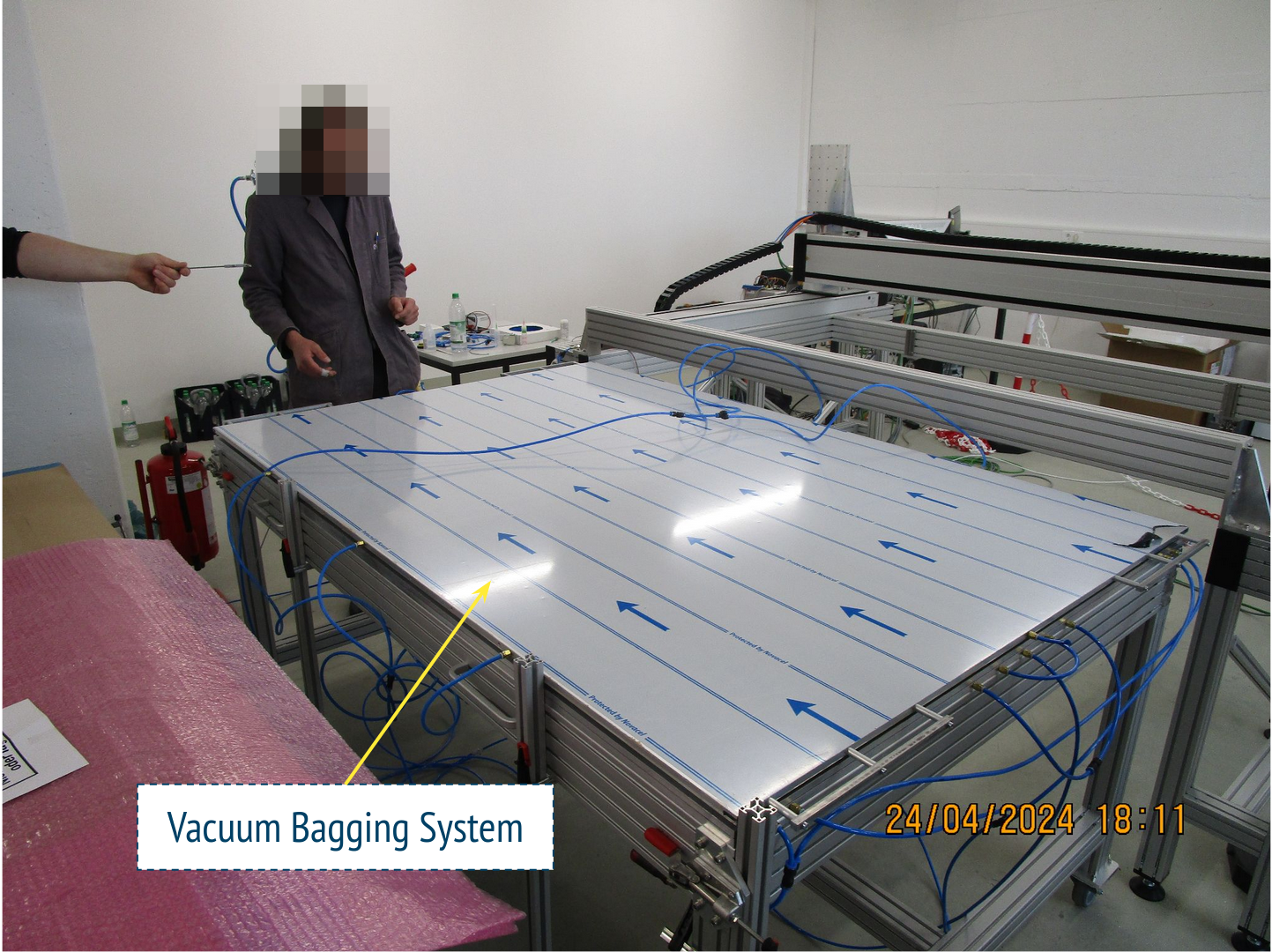} 
        \subcaption{ }
        \label{fig:VacuumBaggingSystemStep1}
    \end{minipage}
    
    \vspace{0.5cm} 
    
    \begin{minipage}{0.48\textwidth}
        \centering
        \includegraphics[width=\textwidth]{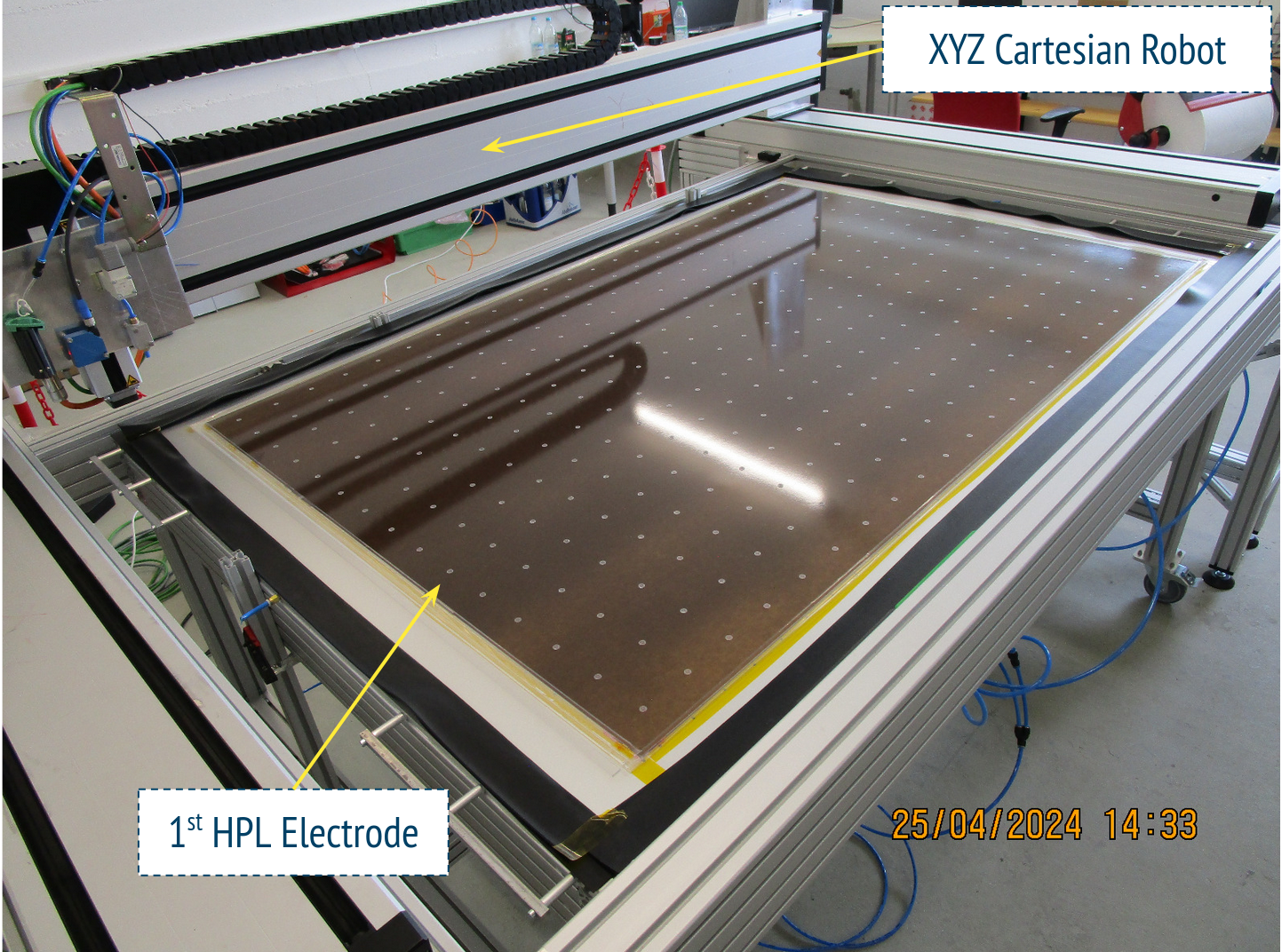} 
        \subcaption{ }
        \label{fig:1stHPLElectrodeFinalization}
    \end{minipage}
    
    \caption{(a) Alignment and positioning of the first HPL electrode plate onto the poly-carbonate spacers and lateral profiles. (b) Vacuum bagging system employed to ensure consistent pressure across the entire assembly during the curing process. (c) Fully assembled HPL electrode with all poly-carbonate components firmly adhered to the surface.}
    \label{fig:GasVolumeAssemblyStep4}
\end{figure}

The first HPL electrode plate is aligned and positioned onto the poly-carbonate spacers and lateral profiles, and the vacuum bagging system is employed to maintain consistent pressure across the assembly, as shown in Figure \ref{fig:1stHPLElectrodeInstallation} and Figure \ref{fig:VacuumBaggingSystemStep1}. This step is critical for achieving uniform adhesion during the curing phase of the epoxy glue. The vacuum bagging system operates at a vacuum level of approximately -200 mbar, corresponding to a force of  $\sim$100 N applied on each spacer. The curing process for the epoxy glue is carried out over a period of approximately 14 hours at a controlled temperature of 20-25$^{\circ}C$ to ensure optimal bonding and structural stability. After the overnight curing phase is complete, the vacuum is shut down, and the first electrode is removed from the vacuum bagging system. The Figure \ref{fig:1stHPLElectrodeFinalization} shows the fully assembled first HPL electrode, with all poly-carbonate components, including the spacers and lateral profiles, firmly adhered to its surface.

\begin{figure}[h!]
    \centering
    \begin{minipage}{0.48\textwidth}
        \centering
        \includegraphics[width=\textwidth]{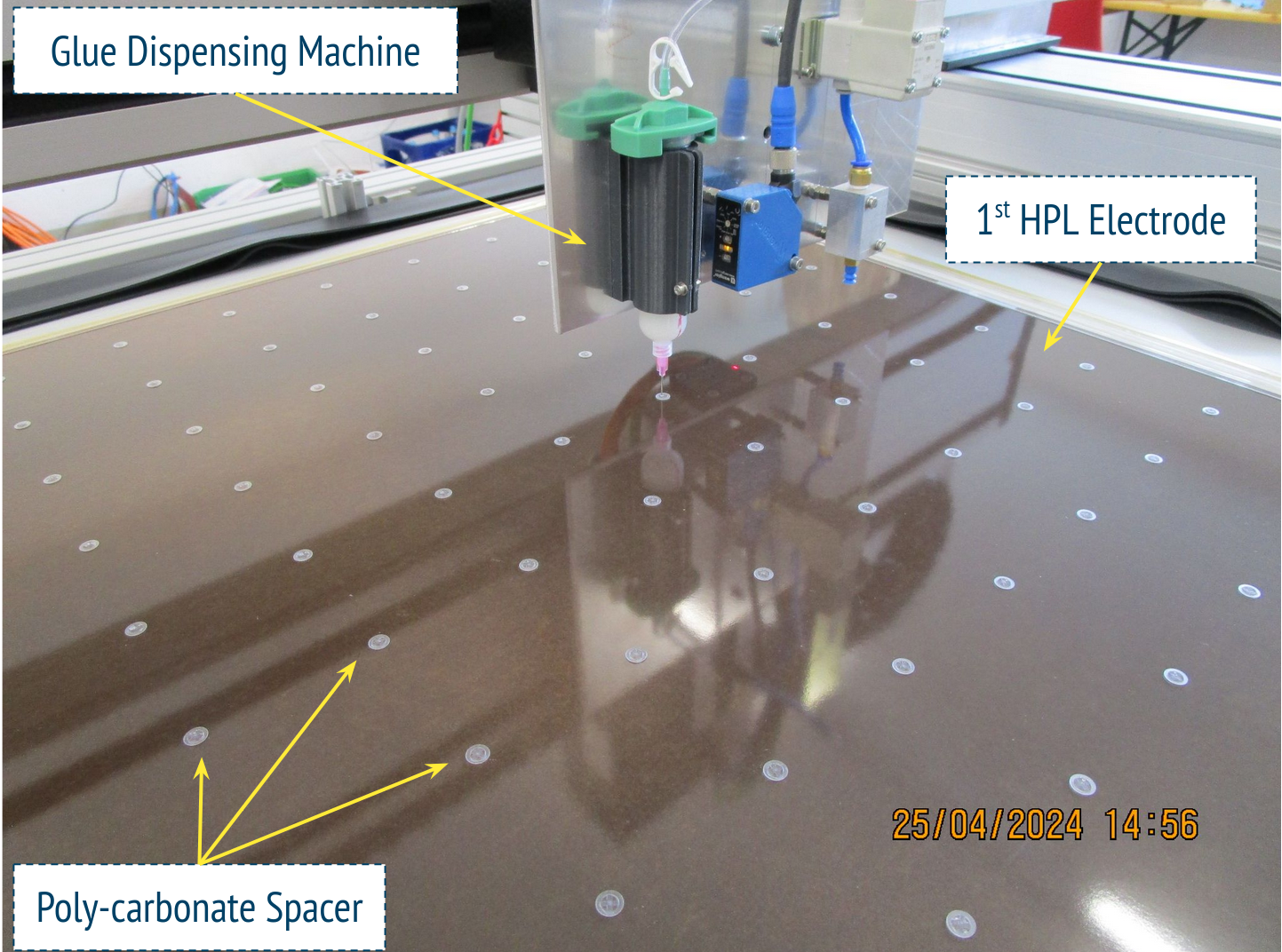} 
        \subcaption{ }
        \label{fig:GluingStep2}
    \end{minipage}
    \hfill
    \begin{minipage}{0.48\textwidth}
        \centering
        \includegraphics[width=\textwidth]{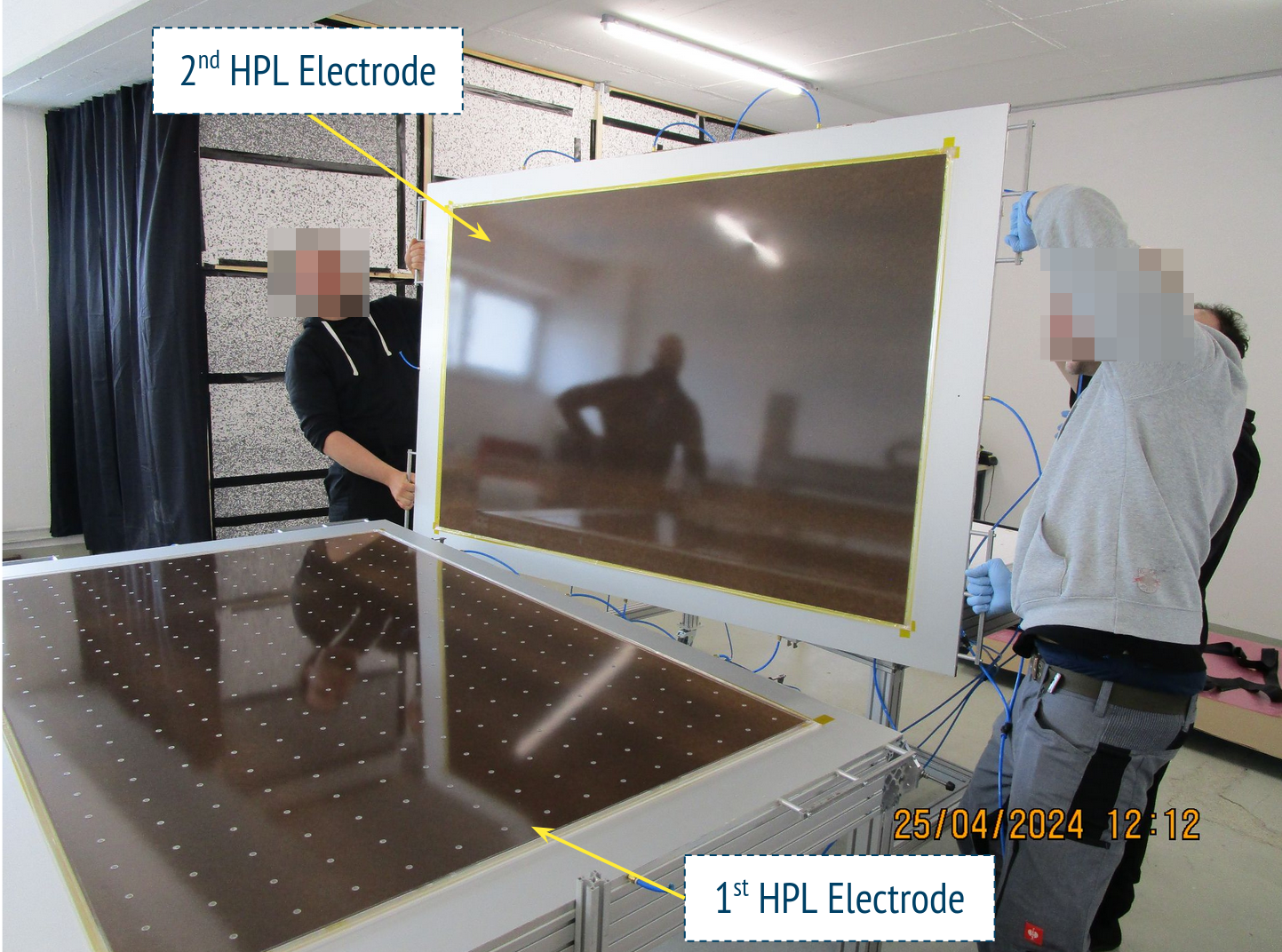} 
        \subcaption{ }
        \label{fig:2ndHPLElectrodeInstallation}
    \end{minipage}
    
    \vspace{0.5cm} 
    
    \begin{minipage}{0.48\textwidth}
        \centering
        \includegraphics[width=\textwidth]{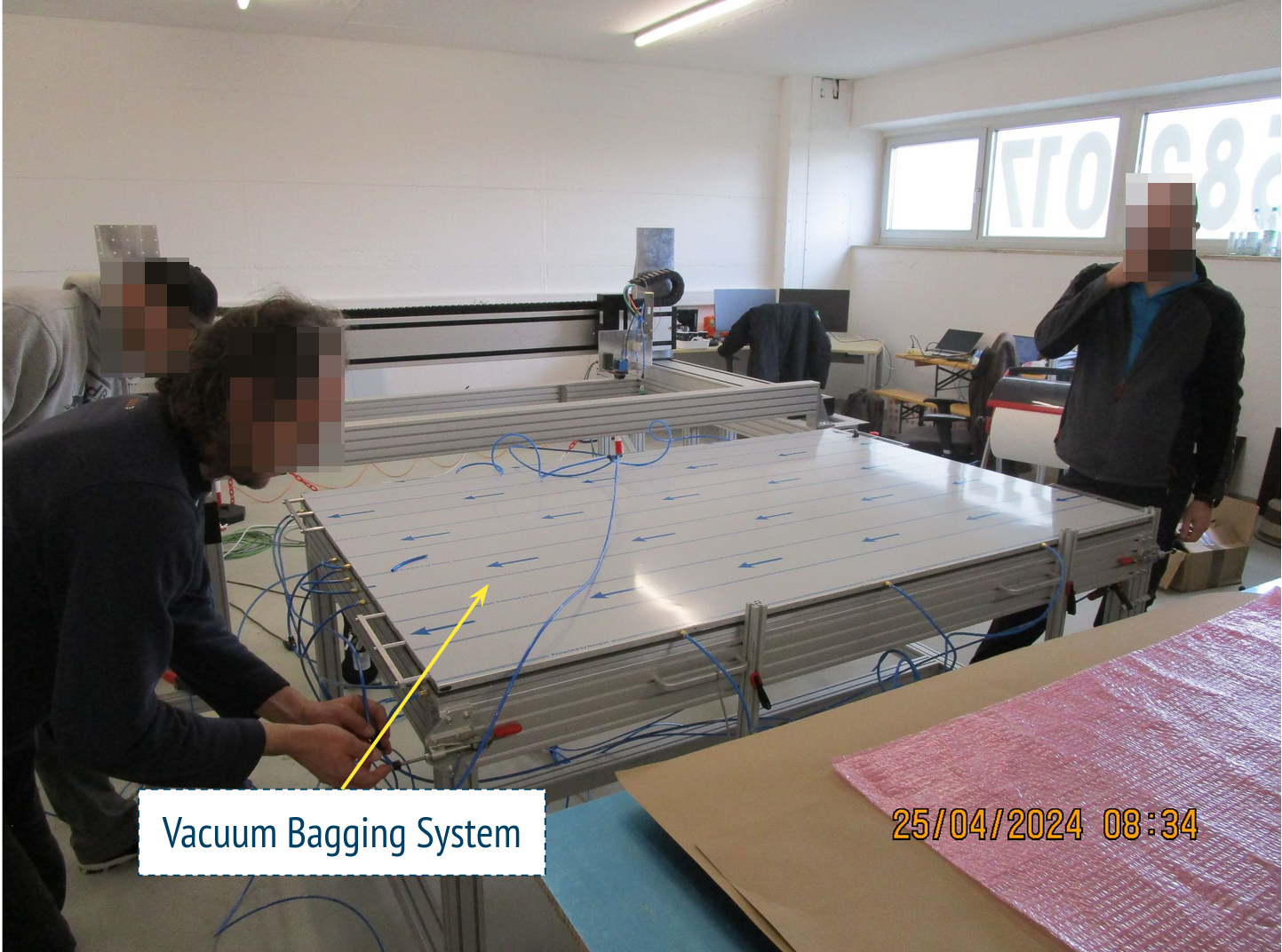} 
        \subcaption{ }
        \label{fig:VacuumBaggingSystemStep2}
    \end{minipage}
    
    \caption{(a) Dispensing adhesive onto spacers and lateral profiles previously affixed to the HPL electrode. (b) Alignment and positioning of the second HPL electrode plate onto the poly-carbonate spacers and lateral profiles. (c) Vacuum bagging system employed to ensure consistent pressure across the entire assembly during the curing process.}
    \label{fig:GasVolumeAssemblyStep5}
\end{figure}

The first HPL electrode plate, with the attached poly-carbonate spacers and lateral profiles, is carefully positioned on the assembly table and secured by a vacuum system. The epoxy glue is applied to the spacers and lateral profiles, following the same protocol as in the preliminary gluing phase, as shown in Figure \ref{fig:GluingStep2}. Subsequently, the second HPL electrode plate is meticulously aligned and placed on top of the spacers and later profiles, as shown in Figure \ref{fig:2ndHPLElectrodeInstallation}. To ensure uniform pressure distribution, the assembly is compressed using a vacuum bagging system and maintained under pressure for a period of 14 hours, allowing sufficient time for the adhesive to fully cure and achieve optimal bonding strength, as shown in Figure \ref{fig:VacuumBaggingSystemStep2}.

\subsubsection{High-Voltage Cable and Gas Pipes Installation}
\label{subsubsec:HVCablesGasPipes}
\begin{figure}[h!]
    \centering
    \begin{minipage}{0.48\textwidth}
        \centering
        \includegraphics[width=\textwidth]{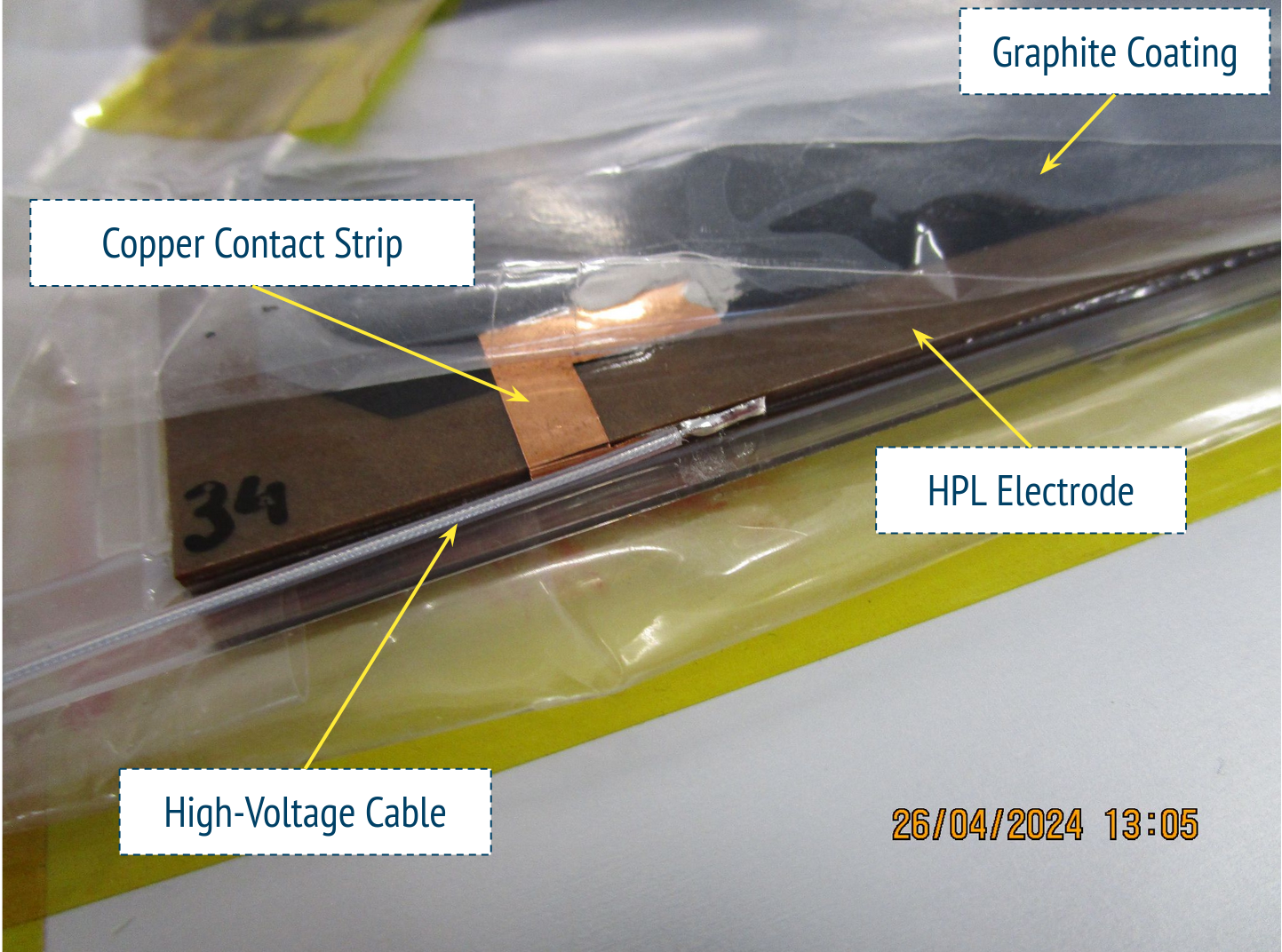} 
        \subcaption{ }
        \label{fig:HighVoltageCableInstallation}
    \end{minipage}
    \hfill
    \begin{minipage}{0.48\textwidth}
        \centering
        \includegraphics[width=\textwidth]{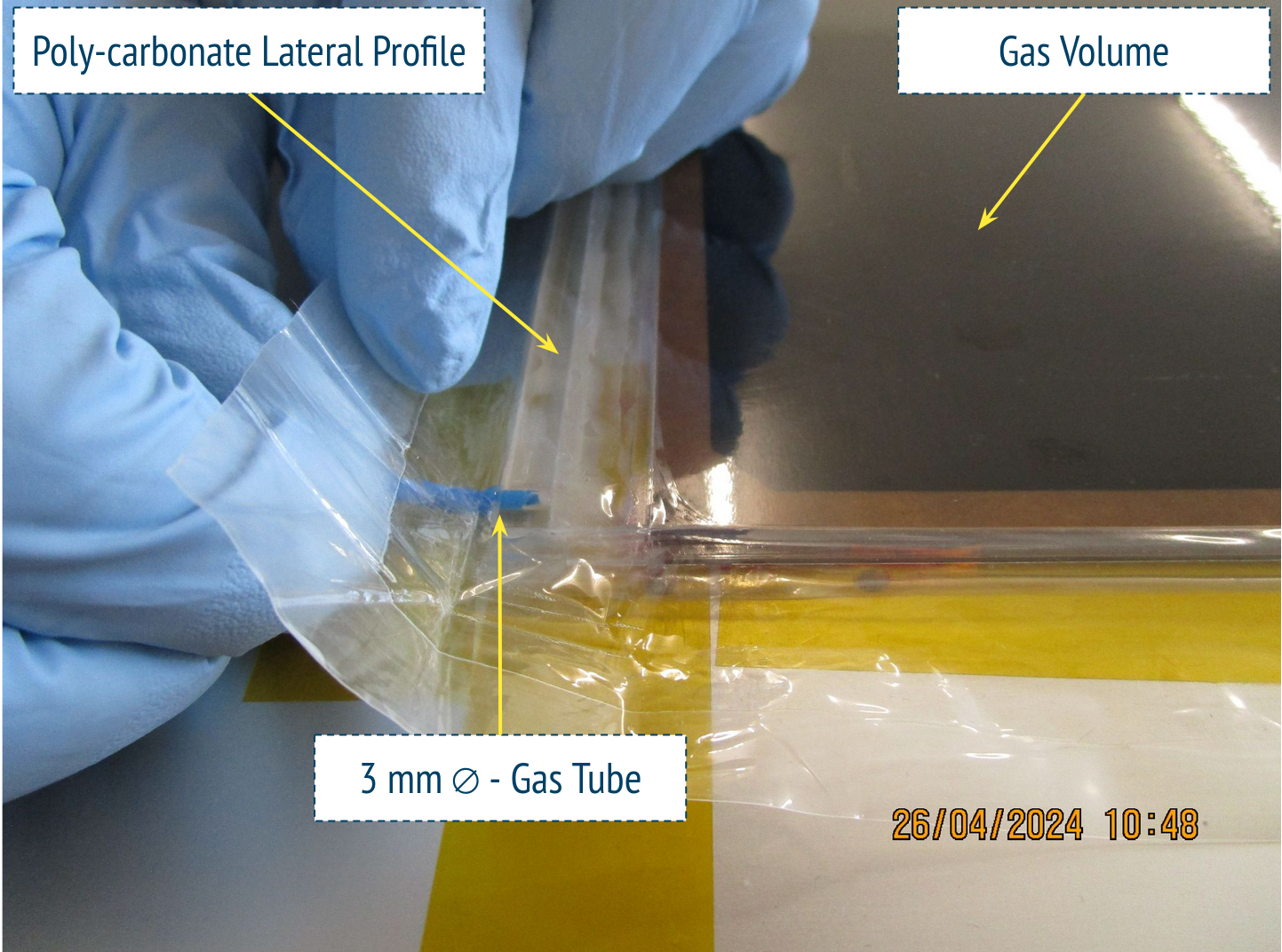} 
        \subcaption{ }
        \label{fig:GasPipeInstallation}
    \end{minipage} 
    \caption{(a) Installation of the high-voltage cable soldered to the copper contact strip on the graphite-coated electrode. (b) Gas tube installation through openings in the lateral profile. Gas tubes are positioned at each corner for uniform gas distribution.}
    \label{fig:GasVolumeAssemblyStep6}
\end{figure}

Following the gluing of the gas volume, an 18 kV rated high-voltage cable is soldered to the copper contact strip, which has been previously affixed to the graphite-coated surface of the electrodes using a conductive silver adhesive, as shown in \ref{fig:HighVoltageCableInstallation}. The gas pipes are connected to the gas volume through openings drilled into a lateral profile glued to the short side of the gas volume, as shown in Figure \ref{fig:GasPipeInstallation}. Each corner is equipped with a gas pipe, and an internal distribution system has been designed to ensure uniform gas flow across the active volume of the detector. 

\subsubsection{Hot-Melt Glue Application and Spacer Indicator Installation}
\label{subsubsec:SpacerIndicators}
\begin{figure}[h!]
    \centering
    \begin{minipage}{0.48\textwidth}
        \centering
        \includegraphics[width=\textwidth]{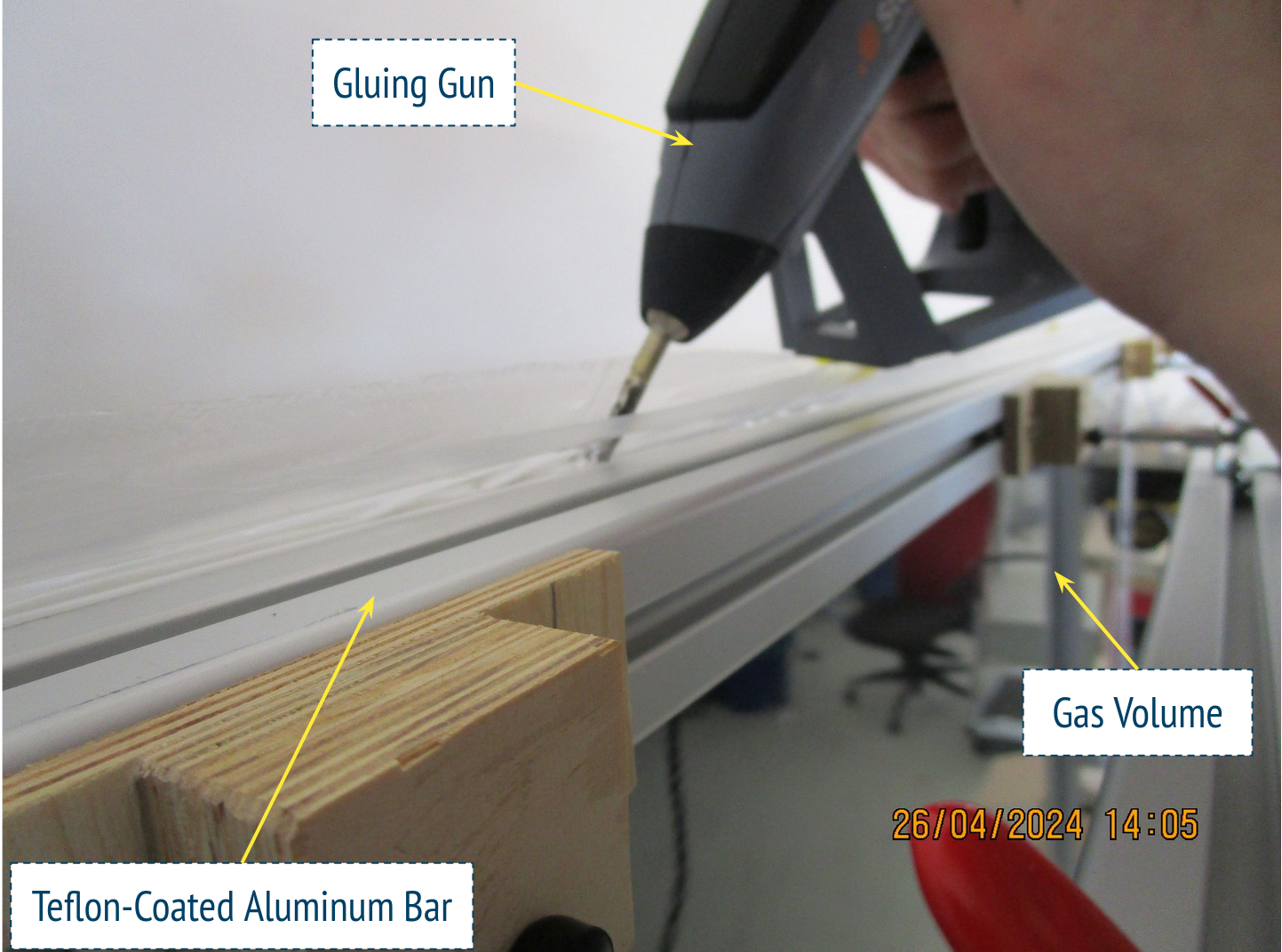} 
        \subcaption{ }
        \label{fig:HotMeltGlueApplication}
    \end{minipage}
    \hfill
    \begin{minipage}{0.48\textwidth}
        \centering
        \includegraphics[width=\textwidth]{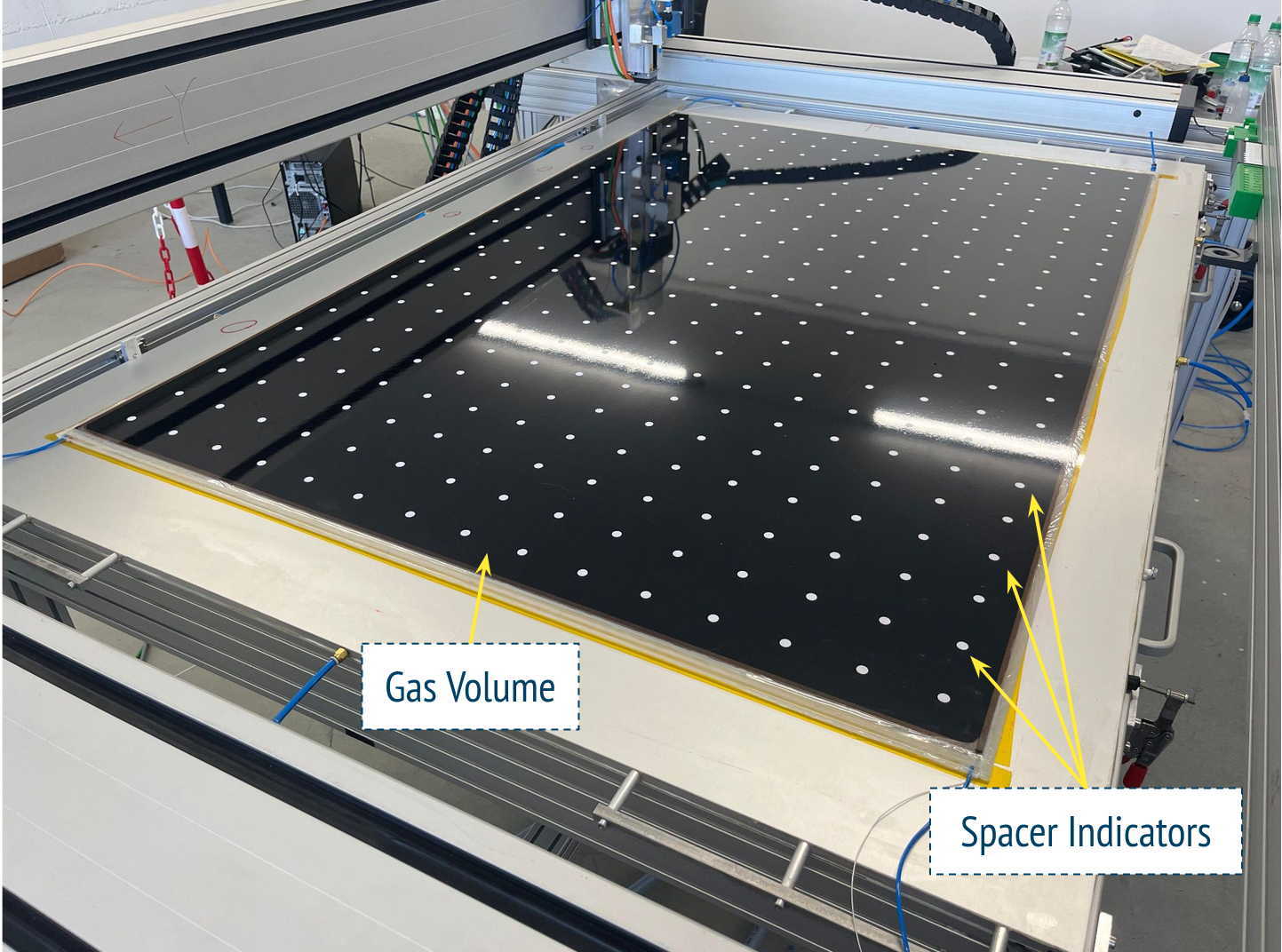} 
        \subcaption{ }
        \label{fig:SpacerIndicatorInstallation}
    \end{minipage} 
    \caption{(a) Application of EVA hot-melt glue along the long sides of the gas volume structure. (b) Installation of 0.10 mm-thick paper indicators on gas volume surfaces to precisely determine spacer positions.}
    \label{fig:GasVolumeAssemblyStep7}
\end{figure}

In the final construction process for the gas volume structure, the long sides are filled and sealed using EVA hot-melt glue applied with a gluing gun, as shown in Figure \ref{fig:HotMeltGlueApplication}. Teflon-coated aluminum bars are used during this step to ensure that the hot-melt glue does not increase the thickness of the gas volume along the edges, maintaining the integrity and uniformity of the gap size. The installation of spacer indicators on both surfaces of the RPC gas volume is achieved by applying 0.10 mm-thick paper indicators, which provide supplementary reference points for accurately determining spacer positions, as shown in Figure \ref{fig:SpacerIndicatorInstallation}. The application process is fully automated, ensuring high throughput and precise positioning, which are crucial for maintaining stringent quality standards during the mass production of gas volume.

\subsection{Linseed Oil Surface Treatment}
\label{subsec:LinseedOil}
\begin{figure}[h]
    \centering
    \includegraphics[width=1\textwidth]{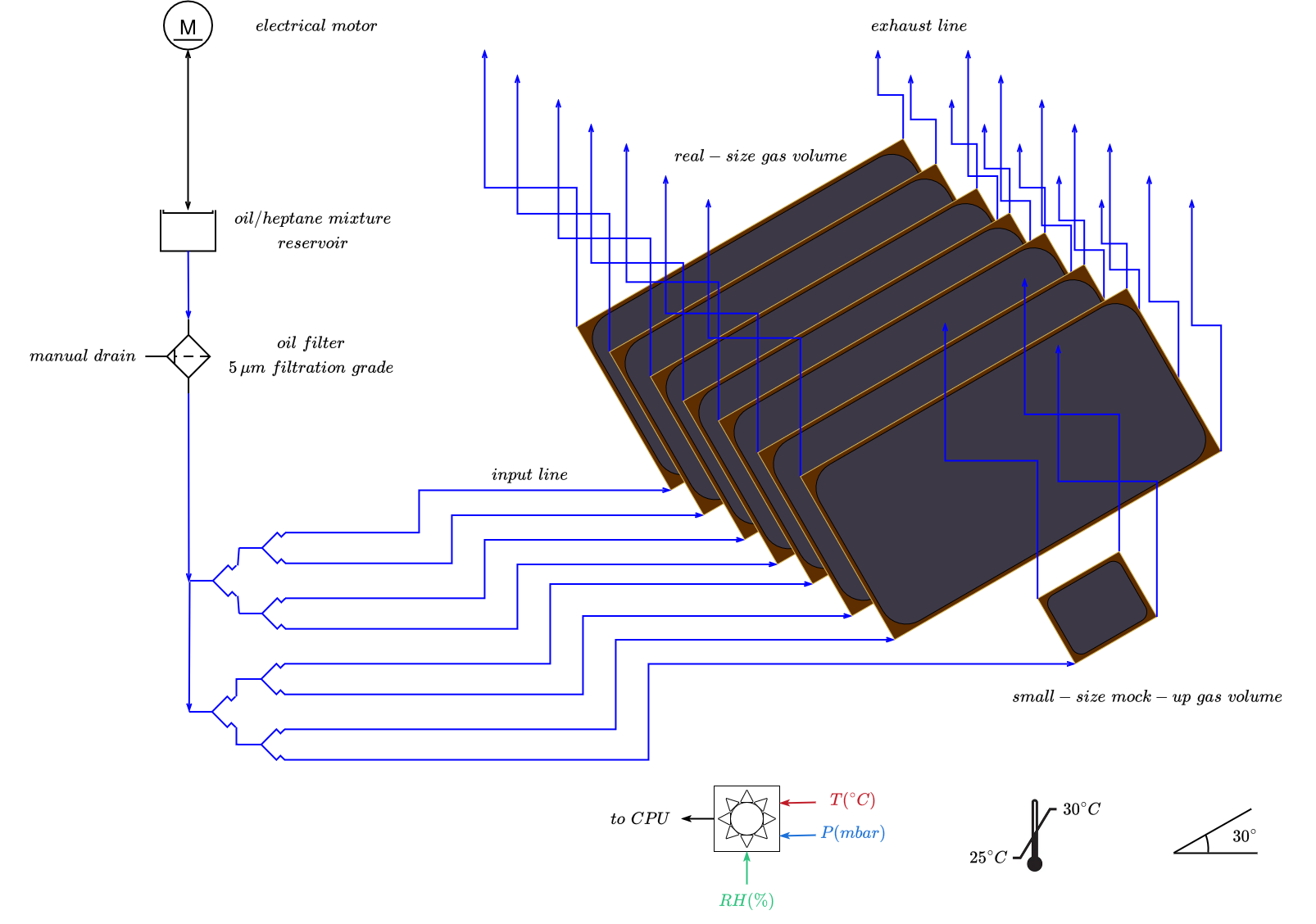} 
    \caption{Diagram of the oiling station for treating the internal surfaces of the gas volume with linseed oil.}
    \label{fig:OilingStation}
\end{figure}

The internal surfaces of HPL electrodes are treated with a thin layer of linseed oil, which significantly enhances the performance and reliability of the detectors. The oil treatment improves the electrode surface quality by creating a smoother and more homogeneous layer, thereby reducing surface irregularities that could lead to localized high electric fields, increased intrinsic current, and elevated noise rates. Furthermore, linseed oil acts as an effective quencher for ultraviolet (UV) photons produced during gas ionization processes, thereby reducing secondary electron emissions and improving the overall stability of the detector. This combination of surface enhancement and photon quenching makes linseed oil treatment a critical step in the fabrication of RPC gas volumes, ensuring optimal detector performance and extended operational lifespan. The standard procedure previously employed by the ATLAS and CMS Collaborations has been refined to achieve optimal results in the linseed oil treatment \cite{2106380}, \cite{Park:2005ze}. Prior to the oiling process, the gas volume undergoes an initial cleaning with pure heptane to eliminate dust, impurities, and potential contaminants introduced during manufacturing. Although strict cleanliness protocols are adhered to throughout production, this additional step ensures the highest level of surface purity, optimizing the internal surface of the gas volume for subsequent oiling processing. 
The precise oiling process begins by filling the gas volume with a meticulously prepared mixture composed of 30\% linseed oil and 70\% heptane. This mixture is introduced via the gas connections from a supply bottle and is carefully filtered through a 5 $\mu m$-rated filter to ensure optimal purity. The linseed oil, designated as GALLOIL and produced by the Italian company GALLO\textsuperscript{\tiny \textregistered}, is a high-purity linseed oil that has undergone a rigorous three-stage heating and purification process to effectively refine the oil, remove impurities and improve its stability and performance under demanding conditions. The heptane is used to reduce the viscosity of the oil, facilitating an even distribution. The operation is carried out in a temperature-controlled environment maintained at 30$^{\circ}C$ to ensure optimal conditions for the coating process. Additionally, in order to prevent blowouts under oil pressure, the gas volume is compressed between two rigid plates, with the entire assembly inclined at an angle of 30$^{\circ}$, as shown in Figure \ref{fig:OilingStation}. 

\begin{figure}[h]
    \centering
    \includegraphics[width=1\textwidth]{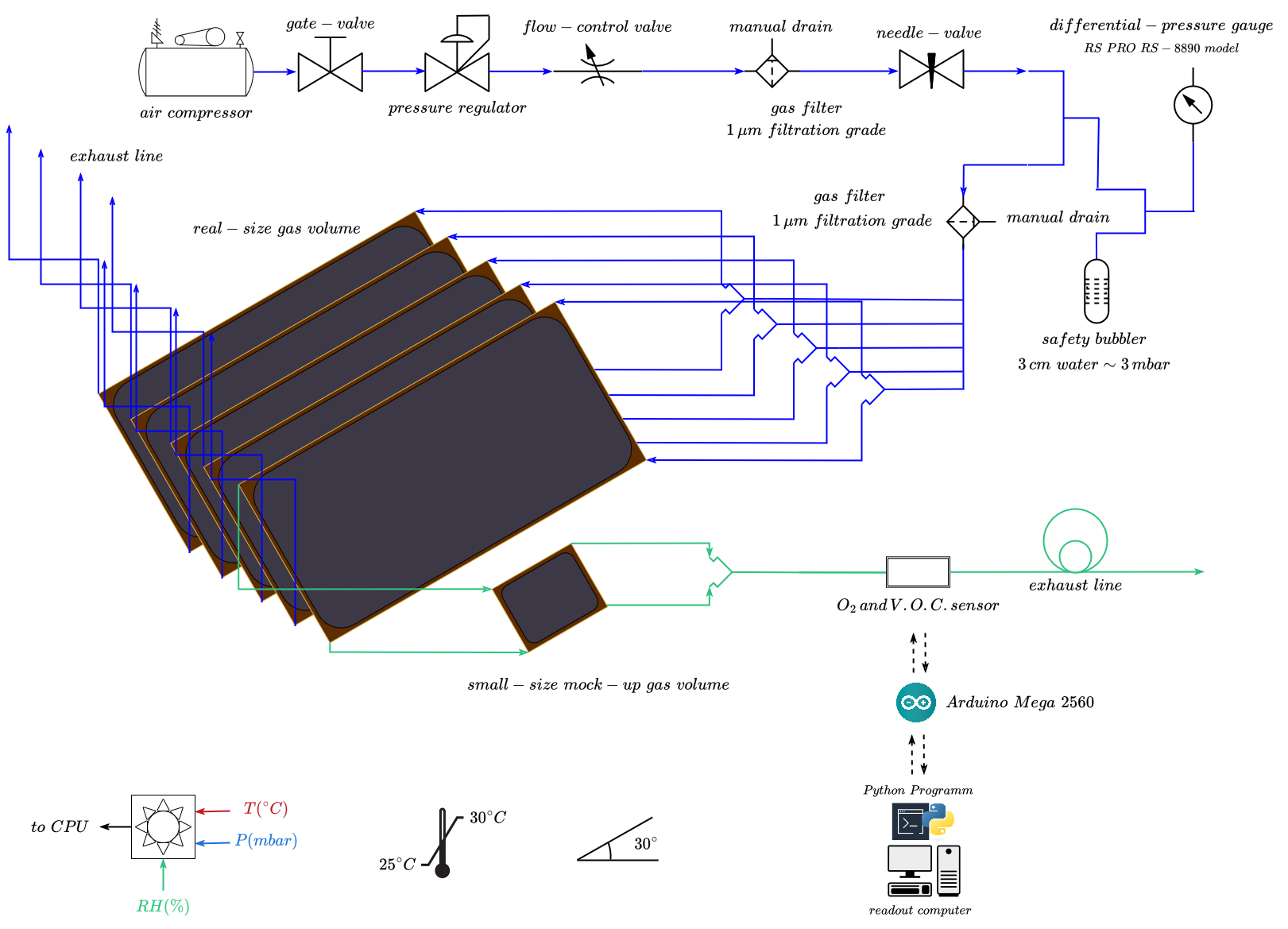} 
    \caption{Diagram of the polymerization station for curing of the linseed oil coating.}
    \label{fig:PolymerizationStation}
\end{figure}

A meticulously designed small-scale gas volume mock-up, faithfully replicating the detector components and materials of the full-scale system, undergoes the same oiling, draining, and drying procedures to rigorously evaluate the quality of the linseed oil coating. Following the filling process, the linseed oil is drained by progressively lowering the supply bottle at a controlled rate of approximately 1 m/h. This gradually draining process is imperative to prevent the formation of droplets and streaks, thereby preserving the uniformity of the linseed oil coating and ensuring the operational reliability of the gas volume. Finally, 1 $\mu$m-grade filtered air is circulated through the gas volume for a duration of two weeks to facilitate the complete polymerization of the linseed oil coating, as shown in Figure \ref{fig:PolymerizationStation}. 

During the initial 24 hours, the flow rate is controlled at 1 l/h to prevent any potential damage to the freshly applied linseed oil layer. Following this initial stabilization period, the flow rate is increased to 2 l/h, accelerating the polymerization process of the linseed oil varnish. This extended conditioning phase ensures thorough polymerization, resulting in a robust and uniform coating on the internal surfaces. 


\begin{figure}[h]
    \centering
    \includegraphics[width=0.5\textwidth]{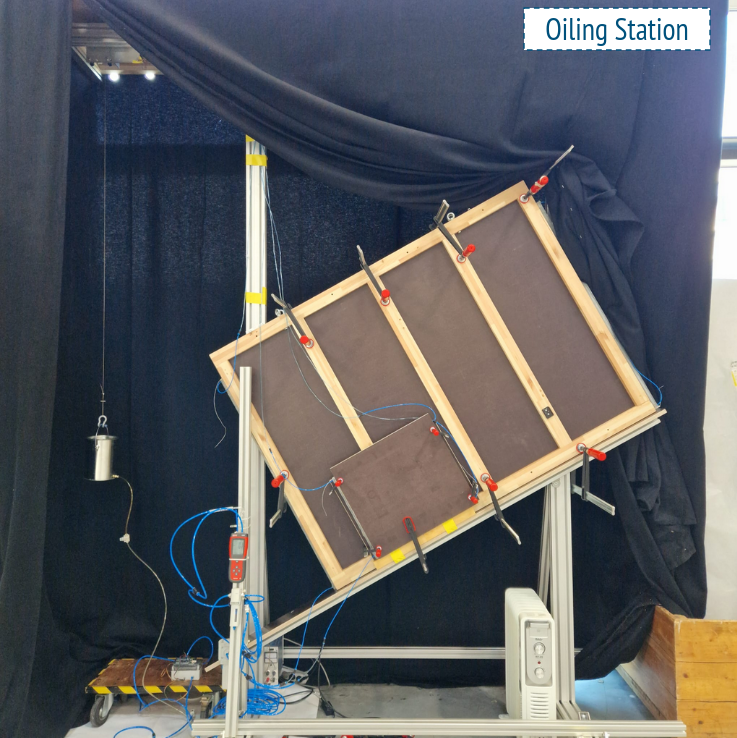} 
    \caption{Oiling station for treating the internal surfaces of the gas volume with linseed oil.}
    \label{fig:OilingStationPic}
\end{figure}


\section{Factory Acceptance Tests}
\label{sec:FactoryAcceptanceTests}
To ensure the highest standards of quality and reliability in RPC gas volume production, a comprehensive Quality Assurance (QA) and Quality Control (QC) program has been established. This program encompasses a series of rigorous protocols and guidelines designed to rigorously monitor and verify every stage of the production process, from material selection and component assembly to final testing and certification.

\subsection{Gas gap mechanical tests}
\label{subsec:GasGapMechanicalTests}
The mechanical tests aim to verify the gas volume tightness and ensure the secure bonding of spacers between the HPL plates. While significant precautions are taken during fabrication, spacers could still detach due to inadequate glue bonding. Additionally, improper handling during assembly can further contribute to spacer detachment, emphasizing the importance of thorough testing to confirm structural integrity.

\subsubsection{Spacer tensile strength measurement}
\label{subsubsec:SpacerTractionMeasurement}
\begin{figure}[h]
    \centering
    \includegraphics[width=1\textwidth]{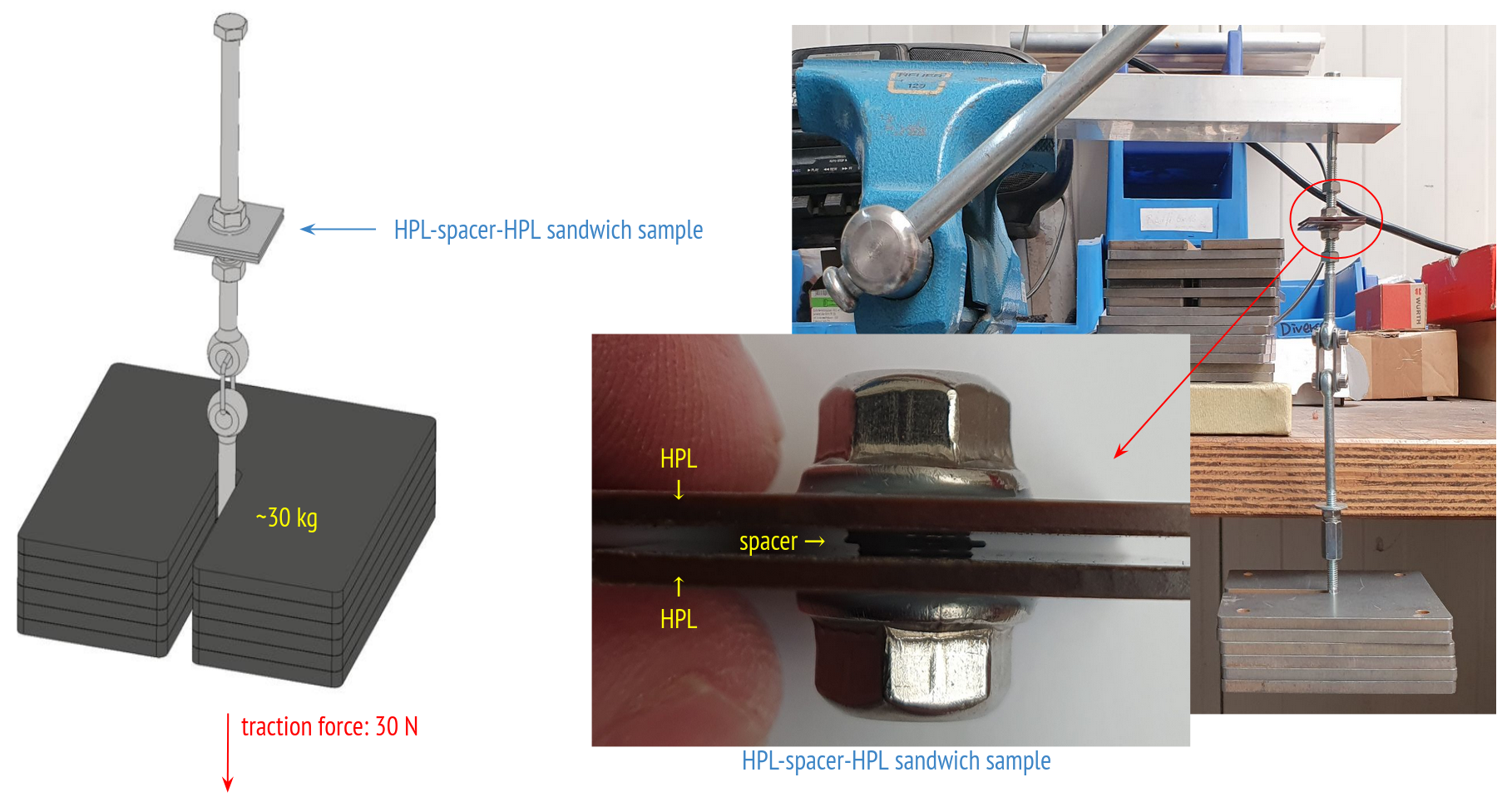} 
    \caption{Schematic 3D model and photograph of the test stand for spacer tensile strength measurement on the $3 \times 3 \, cm^2$ HPL-spacer-HPL sandwich sample.}
    \label{fig:SpacerTractionMeasurement}
\end{figure}

The spacer tensile strength measurement is designed to assess the mechanical reliability of glued pillars on HPL plates, which are essential for ensuring the structural integrity of assembled gas volumes. To conduct this test, a $3 \times 3 \, cm^2$ HPL-spacer-HPL sandwich sample is prepared from each gas volume at the end of each gluing phase. An initial traction force of 30 N is applied to each sample to verify the adhesion strength; if no detachment occurs, the force is then incrementally increased to determine the sample’s breaking point. In standard tests, the breaking point is generally observed at traction forces exceeding 100 N, demonstrating a strong bond between the HPL plates and spacer pillars. This bond strength is critical for maintaining the integrity of the gas volume under operational conditions. Figure  \ref{fig:SpacerTractionMeasurement}  illustrates the test stand employed to assess the tensile strength of spacer pillars, validating the mechanical integrity and robustness of adhesive joints on HPL plates.

\subsubsection{Spacer height measurement}
\label{subsubsec:SpacerHeightMeasurement}
\begin{figure}[h!]
    \centering
    \begin{minipage}{0.48\textwidth}
        \centering
        \includegraphics[width=\textwidth]{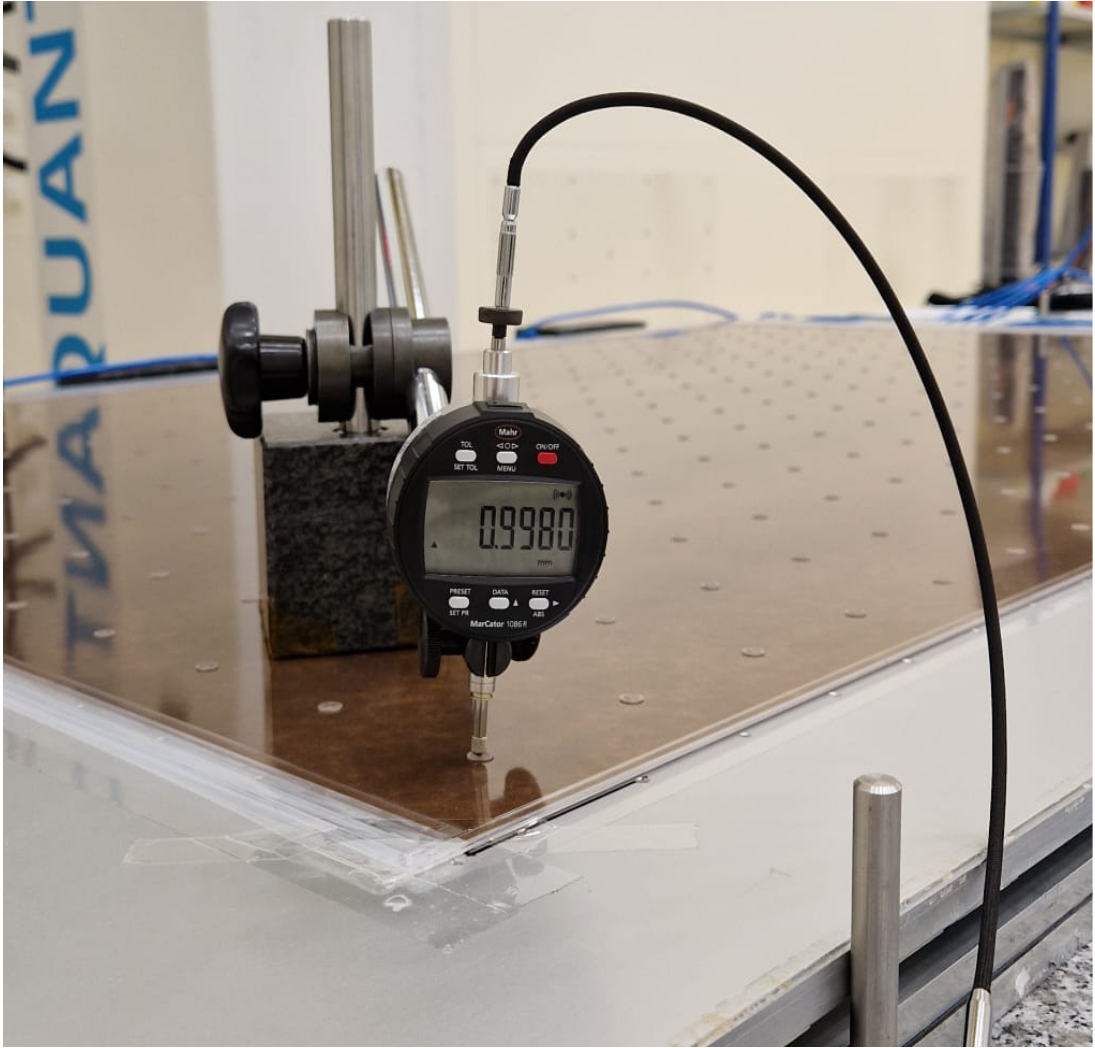} 
        \subcaption{ }
        \label{fig:... ... ...}
    \end{minipage}
    \hfill
    \begin{minipage}{0.48\textwidth}
        \centering
        \includegraphics[width=\textwidth]{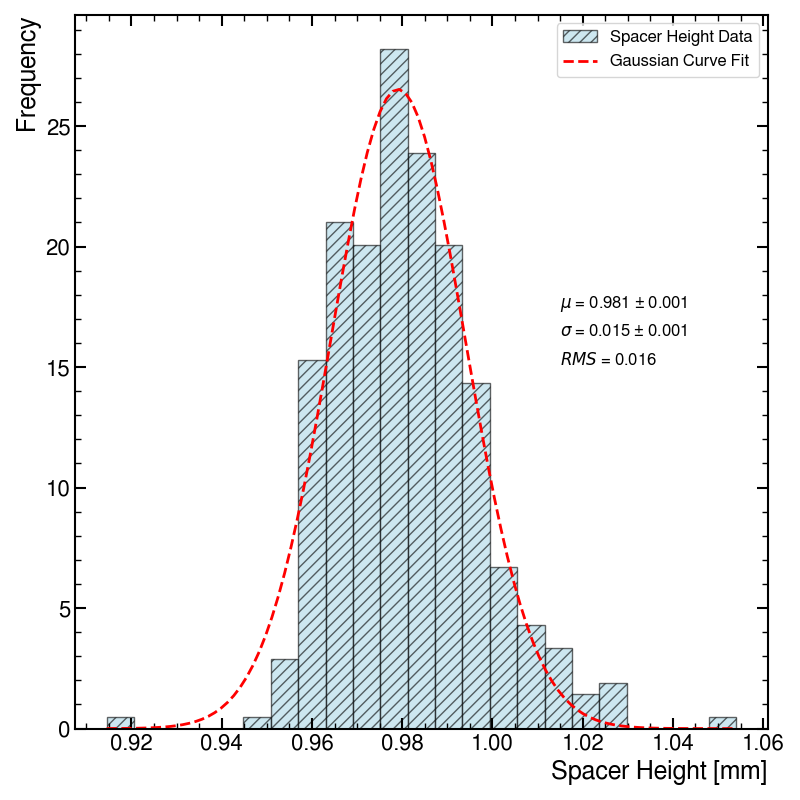} 
        \subcaption{ }
        \label{fig:... ... ...}
    \end{minipage} 
    \caption{Spacer height measurement for a full-scale RPC gas volume prototype. (a) Photograph of the test stand for spacer height measurements. (b) One-dimensional histogram of spacer height measurements. The blue line shows the distribution of the spacer height measurements and the red line is obtained from the gaussian fitting}
    \label{fig:SpacerHeightMeasurement}
\end{figure}

During the gas volume gluing process, maintaining precise spacer height is essential for achieving the required gas volume dimensions, which are critical for optimal detector performance. A quality control step was implemented to verify spacer height as an indirect measure of gas volume accuracy, thereby ensuring the required precision. After the initial adhesive application, the spacers remain accessible, enabling precise height measurements relative to the HPL plates. These measurements are conducted with a high-resolution digital drop indicator, which allows for an accurate assessment of compliance with the 1 mm gas volume specification. Analysis of the measured heights confirms that all spacers meet the gap specification, with deviations confined to a narrow range, remaining well within 20 $\mu$m. This minimal variation highlights the  reliability of the gluing process, ensuring consistent gas volume dimensions throughout the assembly and contributing to the overall stability and reliability of the detector's performance. Representative test results from the spacer height measurement for a full-scale RPC gas volume prototype are presented in Figure \ref{fig:SpacerHeightMeasurement}.

\subsubsection{Gas tightness measurement}
\label{subsubsec:GasLeakTest}
\begin{figure}[h]
    \centering
    \includegraphics[width=1\textwidth]{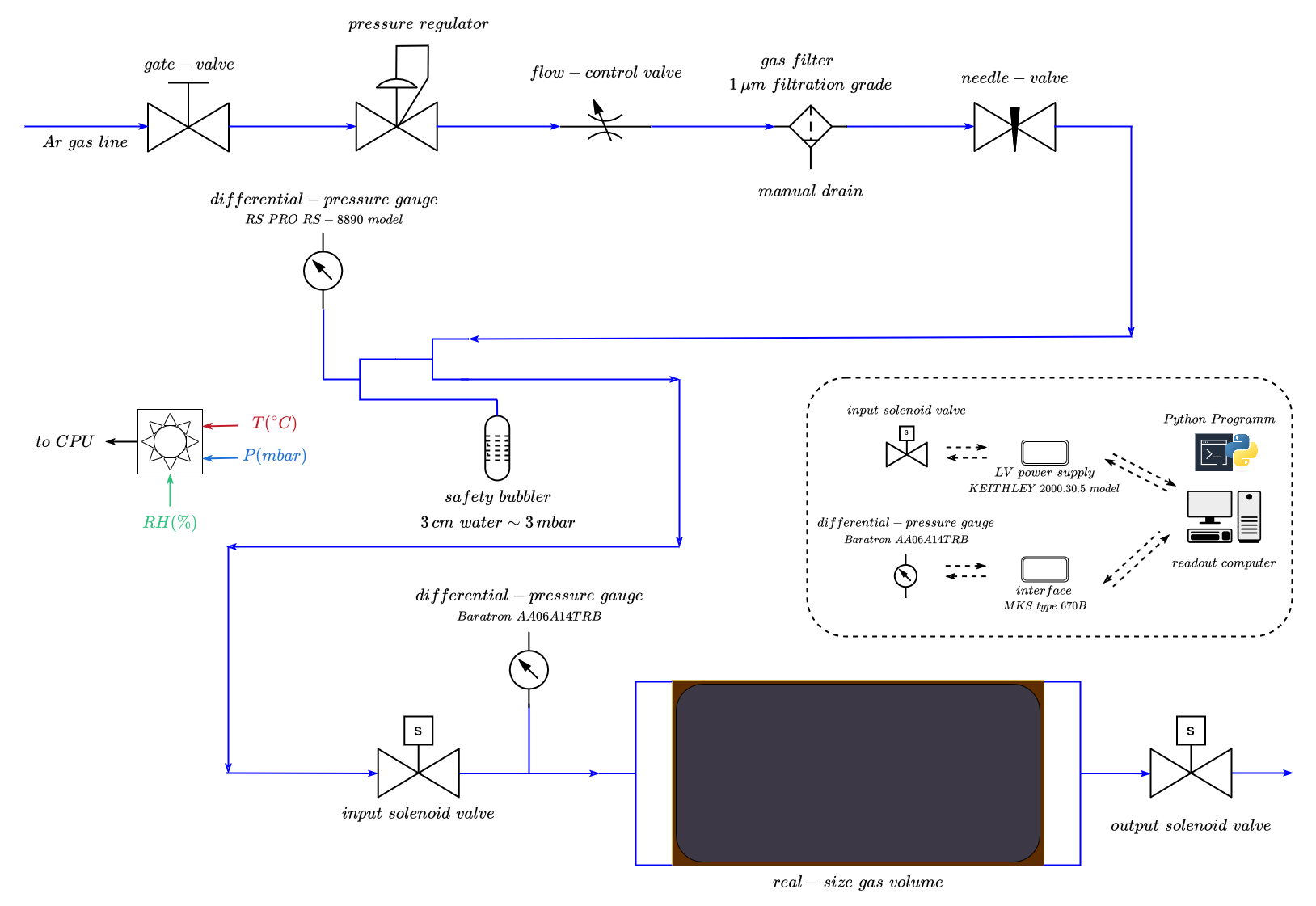} 
    \caption{Diagram of the gas distribution system for leak tightness measurement.}
    \label{fig:SchematicGasSystem}
\end{figure}

The gas tightness test serves as the primary quality control step for newly assembled full-size Resistive Plate Chamber (RPC) gas volumes. This test detects potential leaks and quantifies the leak rate by monitoring the rate of internal over-pressure drop over time. Gas leaks are detrimental, as they waste gas and introduce external contaminants, humidity, and pollutants that can compromise detector performance by triggering reactions in the active gas volume, leading to polymerization and, ultimately, performance degradation. A schematic of the gas system is shown in Figure \ref{fig:SchematicGasSystem}. The setup connects the gas inlet to an argon (Ar) gas source with a pressure regulator. A one-way flow valve at the inlet controls the gas flow rate into the gas volume, while a differential manometer (RS PRO RS-8890) monitors inlet over-pressure. To prevent over-inflation, a safety bubbler vents gas flow at a set threshold (3.0–3.2 mbar above atmospheric pressure). Solenoid valves isolate the detector from the gas system during the measurement, and a high-precision pressure transducer (Baratron AA06A14TRB) records the internal over-pressure. This transducer is connected to a microprocessor-based unit (MKS Type 670B) for power, signal conditioning, and display. Ambient temperature and humidity sensors are positioned near the test stand to account for environmental factors. 

\begin{figure}[h]
    \centering
    \includegraphics[width=0.75\textwidth]{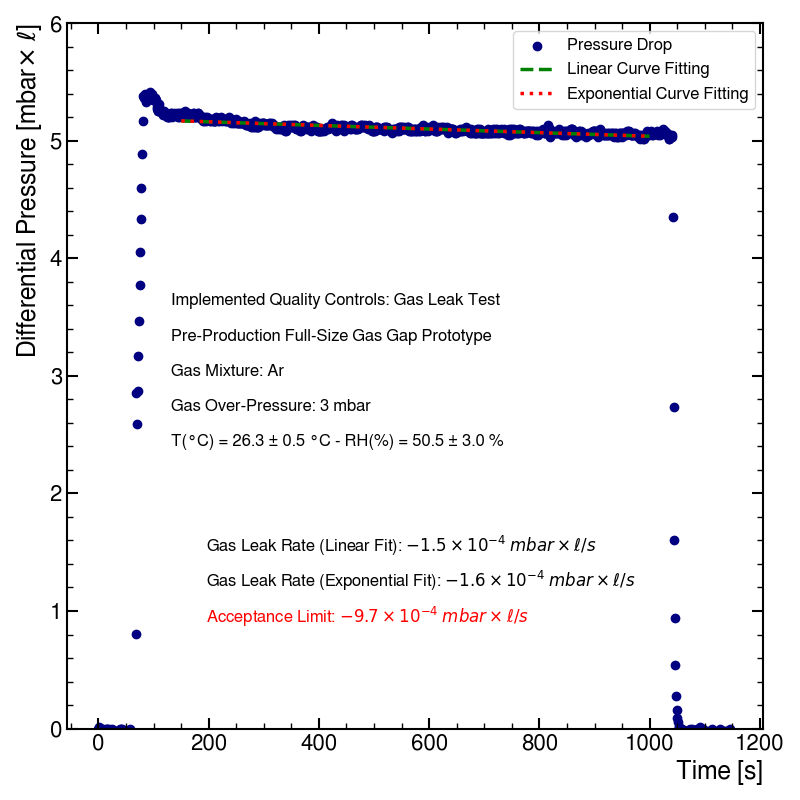} 
    \caption{Results from the gas tightness measurement conducted on a full-size RPC gas volume. The experimental data have been analyzed using both linear and exponential regression models to accurately determine the gas leak rate}
    \label{fig:GasLeakTestResul}
\end{figure}

The test procedure involves pressurizing the gas volume to approximately 3 mbar above atmospheric pressure with pure Ar gas. Pressure is recorded digitally to ensure precision, with QA/QC standards set by the ATLAS Muon Collaboration requiring a pressure loss not exceeding 0.1 mbar over a 3-minute period. After this phase, the gas leak rate is calculated to characterize the pressure decay rate. Once testing is complete, pressure is carefully released to ambient levels. For acceptance, the gas leak rate must be below $9.7 \times 10^{-4} \, mbar \times \ell / s$, corresponding to a 0.1 mbar pressure loss over 3 minutes for a gas volume of approximately 1.74 liters. This stringent criterion ensures long-term reliability and performance of each RPC gas volume in the ATLAS Muon Spectrometer upgrade. Representative test results from the gas tightness  measurement are presented in Figure \ref{fig:GasLeakTestResul}.

\subsubsection{Mechanical rigidity measurement}
\label{subsubsec:MechanicalRigidityMeasurement}
\begin{figure}[h!]
    \centering
    \begin{minipage}{0.70\textwidth}
        \centering
        \includegraphics[width=\textwidth]{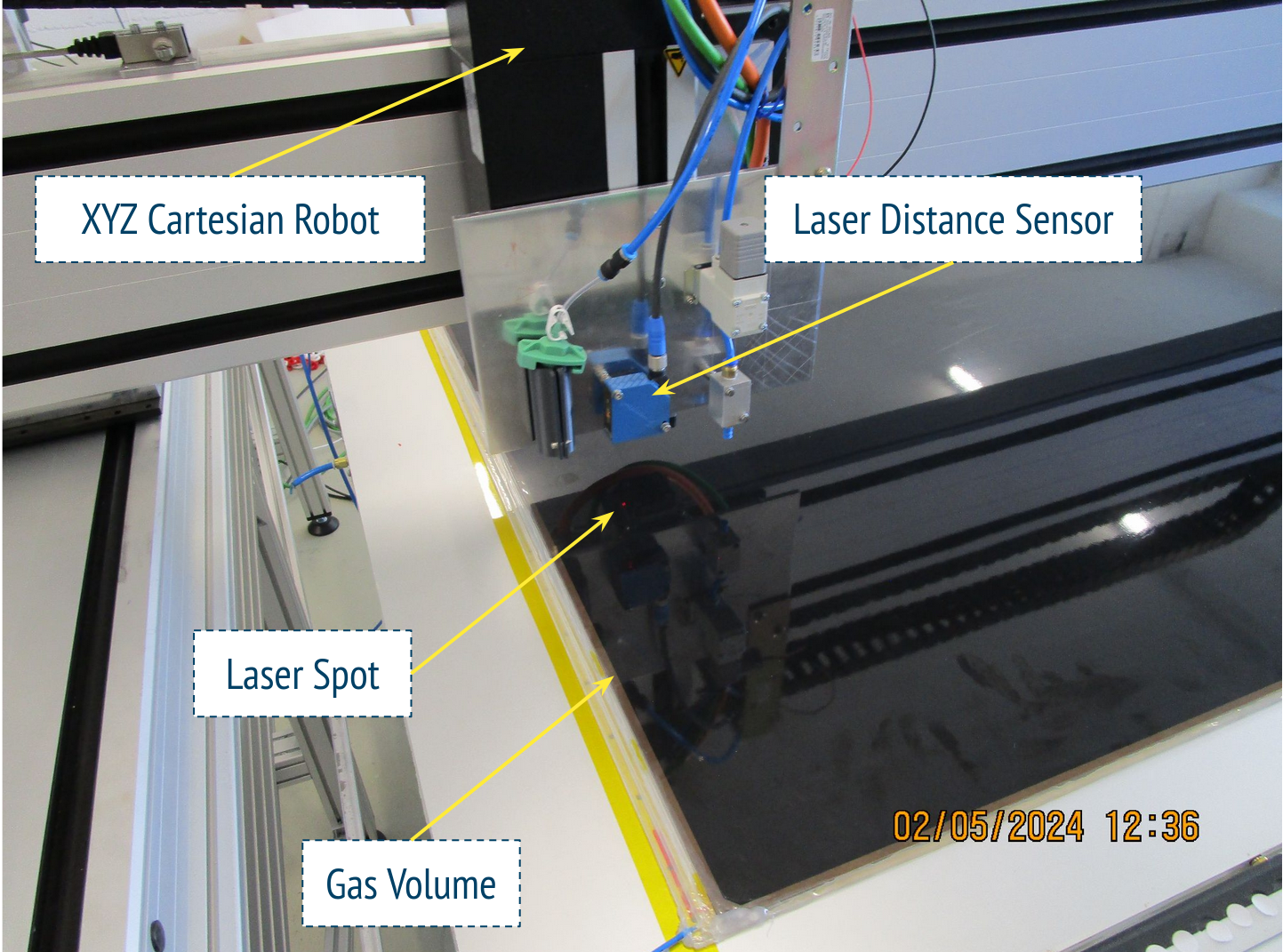} 
        \subcaption{ }
        \label{fig: ... ... ...}
    \end{minipage}
    \hfill
    \begin{minipage}{0.85\textwidth}
        \centering
        \includegraphics[width=\textwidth]{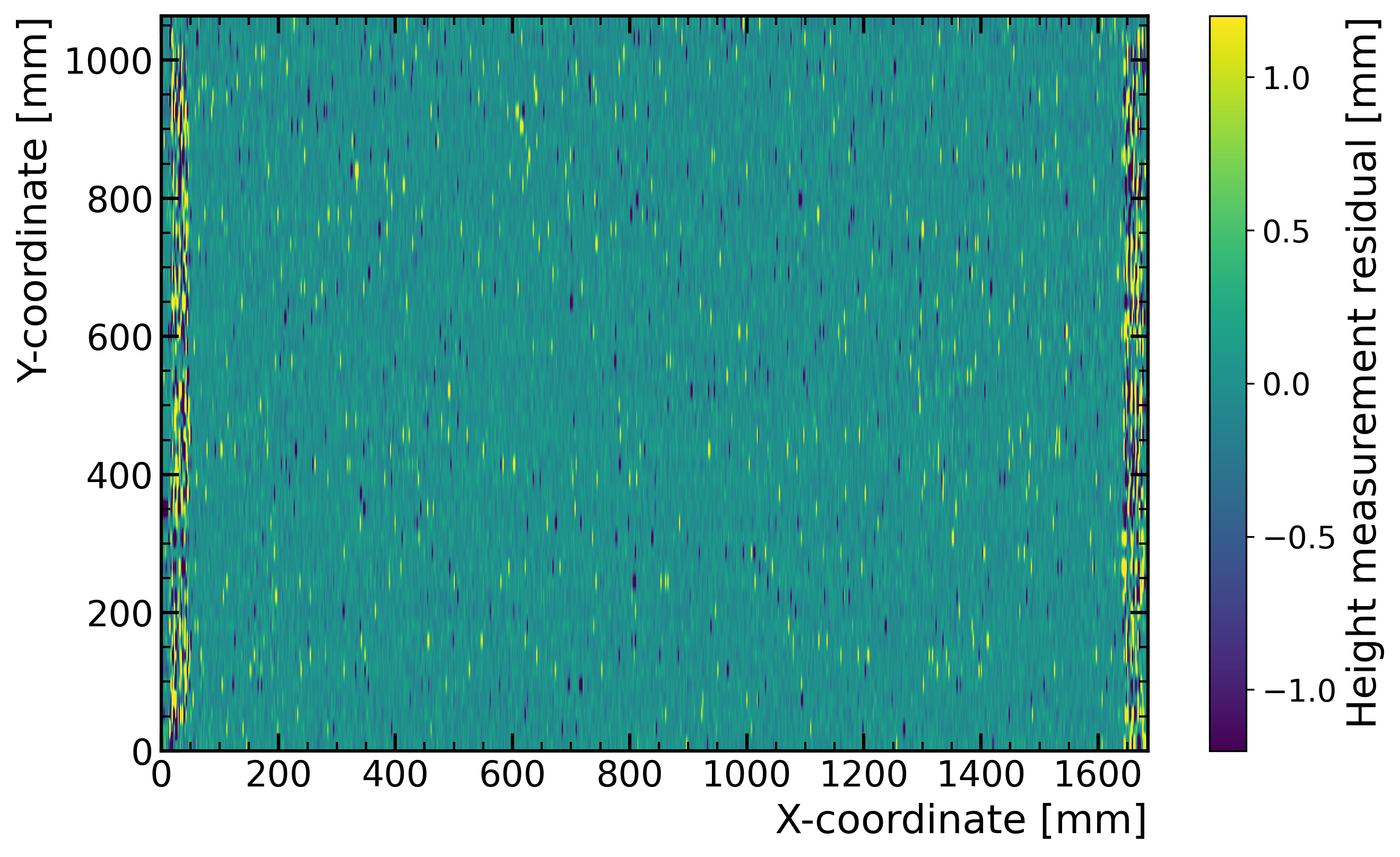} 
        \subcaption{ }
        \label{fig:... ... ...}
    \end{minipage} 
    \caption{(a) Photograph of the test stand for the mechanical rigidity measurement of the gas volume. (b) Two-dimensional histogram of height measurement residuals for the mechanical rigidity assessment of a full-scale gas volume}
    \label{fig:MechanicalRigidityMeasurement}
\end{figure}

Following successful gas leak testing, each full-size RPC gas volume undergoes a mechanical rigidity test to confirm structural integrity. This test verifies the robustness of adhesive bonds and handling quality achieved during fabrication, with a particular focus on detecting any detached spacers. The test utilizes a two-stage laser scanning method. First, a baseline planarity assessment is conducted at atmospheric pressure using a laser scanner, establishing an initial structural profile of the gas volume. Next, the gas volume is subjected to a slight overpressure of approximately 3 mbar, and a second laser scan is performed. This controlled increase in pressure highlights detachment issues that might not be visible at atmospheric pressure. The analysis process involves calculating the discrepancies between the laser scans taken at atmospheric and pressurized conditions. This comparative analysis is essential for identifying any significant changes or deviations that could indicate structural weaknesses.  Through detailed examination of these discrepancies, areas of concern are precisely identified, allowing for corrective measures to reinforce the gas volume’s integrity during the assembly procedure. Representative results from the mechanical rigidity measurement  are presented in Figure \ref{fig:MechanicalRigidityMeasurement}.

\subsection{Leakage current test}
\label{subsec:Leakage current test}
\begin{figure}[h]
    \centering
    \includegraphics[width=1\textwidth]{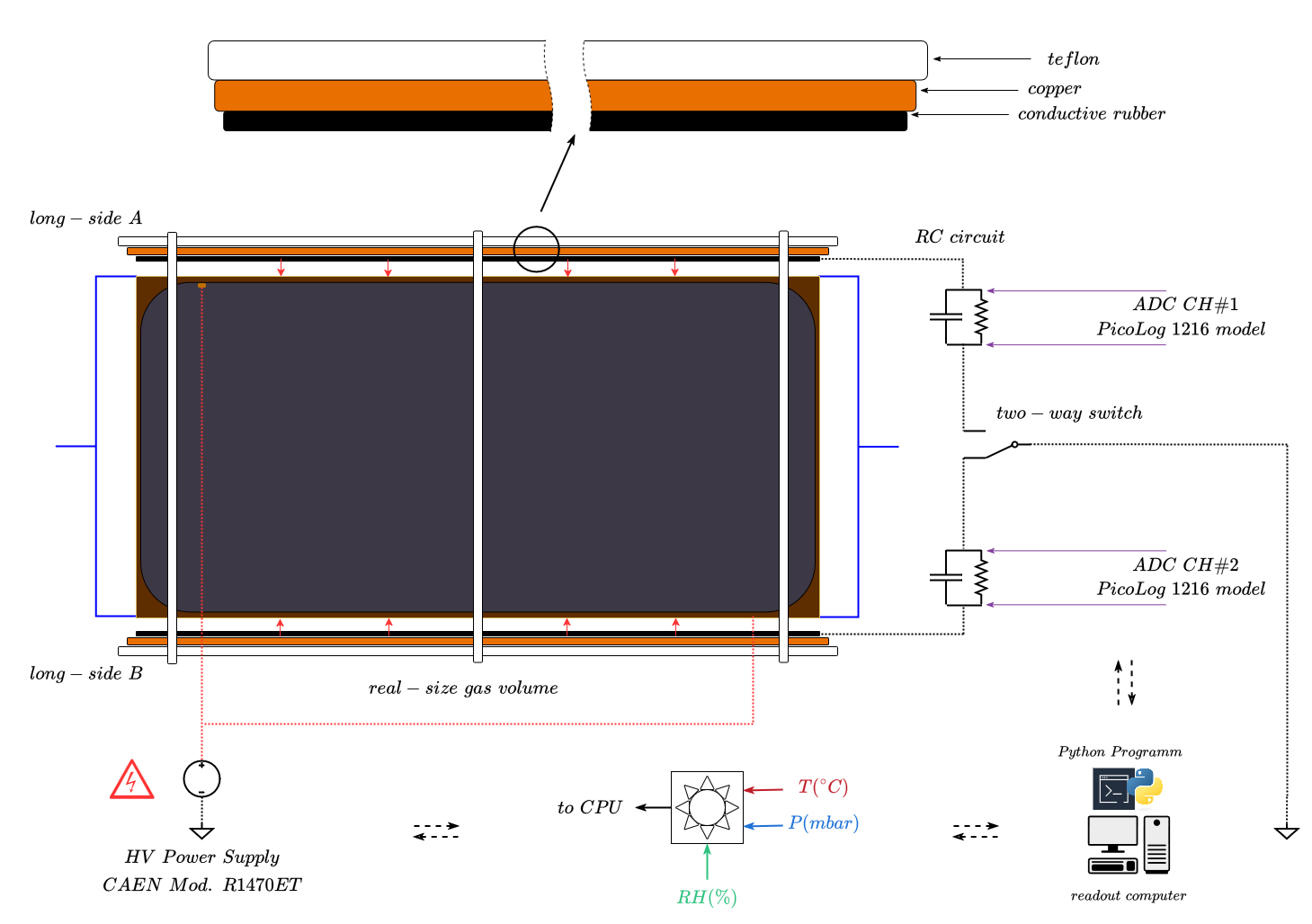} 
    \caption{Diagram of the test stand for leakage current measurement of a full-scale RPC gas volume.}
    \label{fig:LeakageCurrentDiagram}
\end{figure}

The leakage current test serves as a critical quality assurance step to verify the insulation integrity of the High Pressure Laminate (HPL) electrodes within each RPC gas volume. Ensuring proper insulation is essential to both the safety and performance of the RPC detectors. During this test, a high voltage of up to 8 kV is applied to the HPL electrodes. A copper bar equipped with a layer of conductive rubber is mounted on an insulating Teflon support and positioned to ensure uniform contact along the long side of the gas volume. The conductive rubber provides a compliant electrical interface, while the copper bar is connected to ground through a resistor in order to allow the measurement of any leakage current.
The leakage current is determined by measuring the voltage drop across the resistor via an Analog-to-Digital Converter (ADC), providing a sensitive indication of current leakage, as shown in Figure \ref{fig:LeakageCurrentDiagram}. The measurement procedure is fully automated and sequentially performed on both long sides of the gas volume. The test is first conducted on one side (referred to as long-side A), after which the system automatically switches to the opposite side (long-side B) to repeat the measurement. During the measurement on one side, the other side is electrically disconnected by a two-way switching device, ensuring well-defined test conditions and eliminating the risk of parasitic current paths through the inactive electrode.

\begin{figure}[h]
    \centering
    \includegraphics[width=0.75\textwidth]{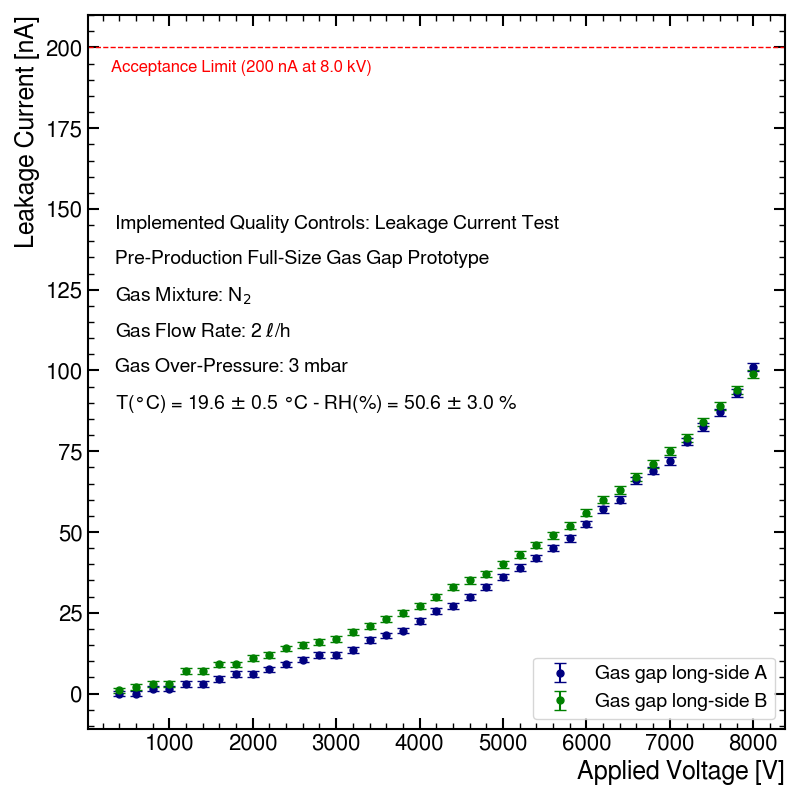} 
    \caption{Leakage current test for a full-scale RPC gas volume}
    \label{fig:LeakageCurrentTest}
\end{figure}

To comply with quality standards, the leakage current must remain below 200 nA at maximum applied voltage. This strict threshold ensures robust insulation quality of each gas volume, minimizing the risk of electrical failures that could compromise detector reliability and safety. Representative results from the leakage current test are presented in Figure \ref{fig:LeakageCurrentTest}.

\subsection{Linseed oil coating quality‑control tests}
\label{subsec:LinseedOilQualityControl}
Quality-control tests for linseed oil coatings are essential to guarantee durable and high-performance protective layers within RPC gas volumes. A comprehensive validation approach involves two key methodologies. Firstly, real-time monitoring of the polymerization dynamics of linseed oil enables precise characterization of the curing process and ensuring optimal polymerization conditions. Secondly, a dedicated mock-up gas volume is opened upon completion of the polymerization phase for meticulous visual inspections. This step provides direct assessment of coating uniformity, adhesion quality, and the identification of defects, ensuring that the linseed oil coating adheres strictly to rigorous quality standards required for stable, long-term detector performance.

\subsubsection{Real‑time monitoring of linseed oil polymerization dynamics}
\begin{figure}[h]
    \centering
    \includegraphics[width=0.7\textwidth]{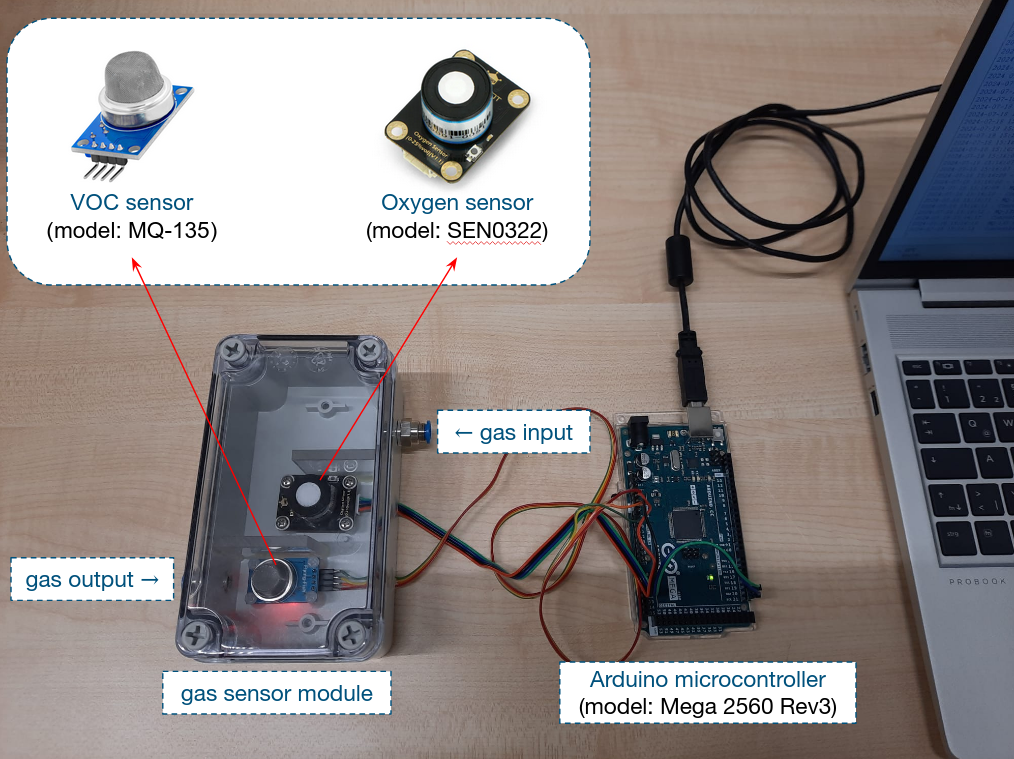} 
    \caption{Internal view of the gas sensor module showing the VOC sensor and the oxygen sensor. The module is connected to an Arduino micro-controller for data acquisition and processing.}
    \label{fig:GasSensorModule}
\end{figure}

Real‑time tracking of linseed‑oil polymerisation relies on the commercial MQ‑135 sensor manufactured by Winsen Electronics Technology Co.; its nanocrystalline n‑type tin‑dioxide ($SnO_2$) sensing layer is tailored for high‑sensitivity detection of a broad spectrum of volatile organic compounds (VOCs). The semiconductor-based VOC monitoring system is mounted on the exhaust line of a mock-up gas volume downstream of the drying airflow after the linseed oil treatment and interfaced with an Arduino micro‑controller for continuous data acquisition, as shown in Figure \ref{fig:PolymerizationStation}. In order to prevent premature degradation of the sensing layer by heptane and unreacted linseed‑oil vapours, the sensor module is connected only from the second week of curing, when the effluent is dominated by oxidative cross‑linking by‑products. In clean air $SnO_2$ exhibits a comparatively low charge‑carrier density and therefore a high electrical resistance; adsorption of VOC gases (e.g., aldehydes, ketones, alcohols) donates electrons to the conduction band, markedly increasing conductivity and correspondingly lowering the sensor resistance. The purity of the filtered, compressed air that is continuously purged through the gas volume, both to supply molecular oxygen for autoxidative cross‑linking and to remove volatile by‑products, is monitored by the dimensionless sensor response:
\begin{equation}
\chi_{air} = \frac{R_S}{R_0}
\label{eq:airquality}
\end{equation}
\noindent where $R_S$ is the instantaneous resistance of the $SnO_2$ sensing layer in the circulating stream and $R_0$ is the baseline resistance established in clean reference air. A deviation of the sensor response $\chi_{air}$ below unity therefore signals an accumulation of reducing VOCs within the gas volume, whereas values approaching unity certify that the flushing air has regained its original degree of cleanliness and that polymerisation is nearing completion. Furthermore, the oxygen molar fraction in the purge stream is logged with a galvanic electrochemical probe (Gravity $O_2$ sensor, model SEN0322, DFRobot), interfaced via the same $I^2C$ bus that serves the VOC array. Continuous read‑out ensures that the partial pressure of $O_2$ remains at the ambient benchmark of $\approx 21 \, vol\%$, thereby securing the oxidant reservoir required for the free‑radical autoxidative chain mechanism governing linseed‑oil cross‑linking. Figure \ref{fig:GasSensorModule} shows the VOC sensor and the oxygen sensor housed within the gas sensor module, which is interfaced with an Arduino micro-controller.

\begin{figure}[h]
    \centering
    \includegraphics[width=0.75\textwidth]{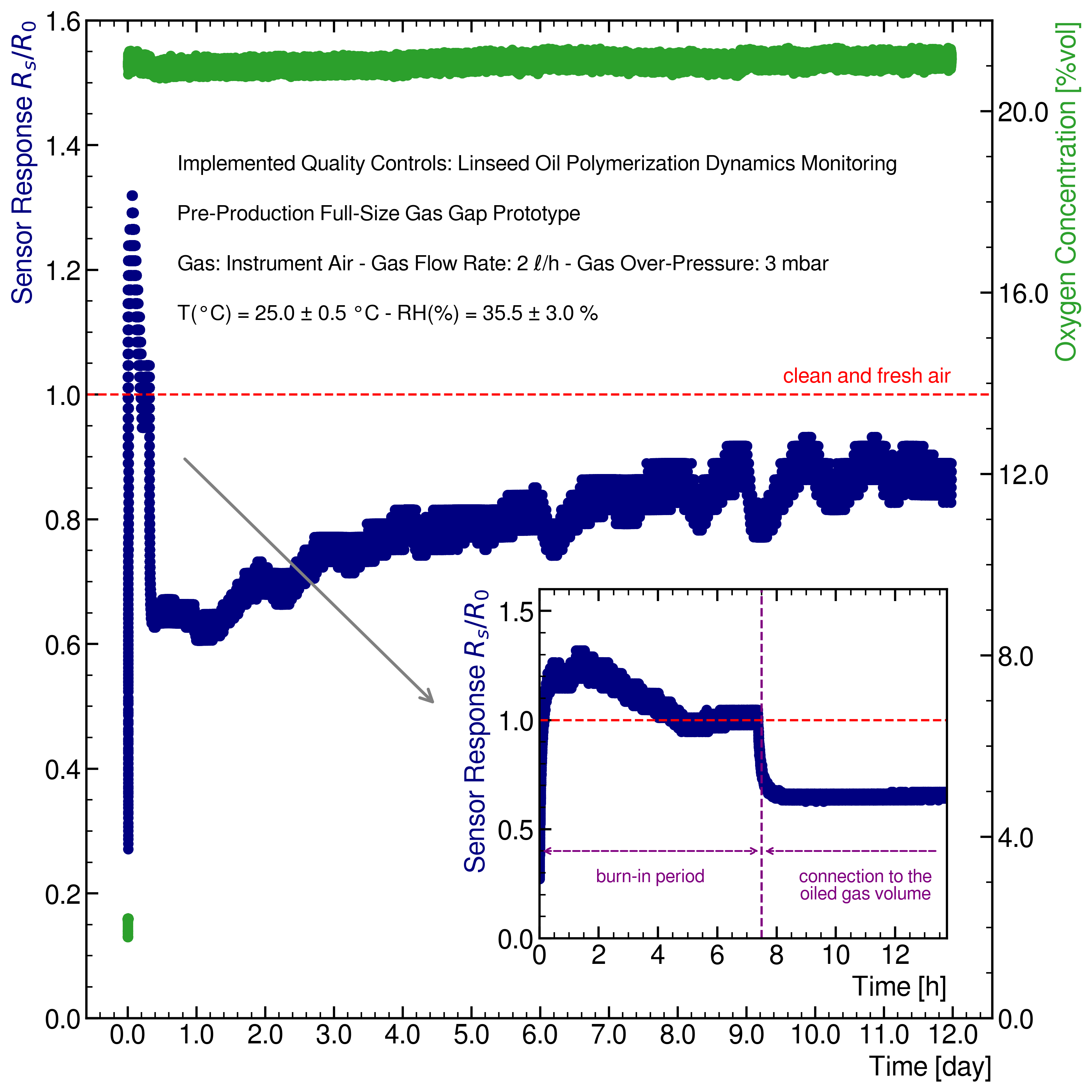} 
    \caption{Real-time monitoring of linseed oil polymerization dynamics after oiling treatment in a full-scale RPC gas volume via a VCO sensor and an oxygen sensor.}
    \label{fig:VOCsensorplots}
\end{figure}

The inset in Figure \ref{fig:VOCsensorplots} presents the early‑stage conditioning of the VOC probe, plotting the dimensionless sensor response $\chi_{air}$ (left ordinate) alongside the oxygen molar fraction (right ordinate) during the first few hours after after power‑up the sensor module, before attachment to the exhaust manifold. A brief burn‑in interval is evident, during which $\chi_{air}$ overshoots to values $>1.2$ because the $SnO_2$ surface acquires adsorbed oxygen species and the heater reaches thermal steady state. Once stabilized, a calibration phase in clean, filtered compressed air brings $\chi_{air}$ to unity while oxygen molar fraction remains constant at $\approx 20.9 \, vol\%$, confirming an uncontaminated environment. After the calibration phase the sensor module is coupled to the exhaust line of the oiled gas volume; the sensor response $\chi_{air}$ plummets below 0.70 within minutes, indicative of the sudden appearance of reducing VOCs that donate conduction‑band electrons to the $SnO_2$ lattice. The oxygen reading is unperturbed, demonstrating that the VOC burst arises from chemical evolution of the coating rather than dilution of the purge stream. The Figure \ref{fig:VOCsensorplots} prolongs the observation window to the second week of polymerization. After the initial drop, the sensor response $\chi_{air}$ rises monotonically, approaching an asymptote of 0.85–0.90 by the end of the second week of polymerization. The gradual recovery of the sensor response toward the clean‑air reference demonstrates that, as polymerization proceeds, the emission rate in reducing VOCs diminishes and the atmosphere in the gas volume reverts to near‑ambient quality. The convergence of the sensor response toward unity therefore furnishes a self‑consistent endpoint for terminating the curing stage, ensuring a complete cross‑linking and preserving film integrity.

\subsubsection{Mock-up gas volume opening and subsequent visual inspection}
\begin{figure}[htbp]
  \centering
\begin{subfigure}[b]{.49\textwidth}
  \includegraphics[width=\linewidth]{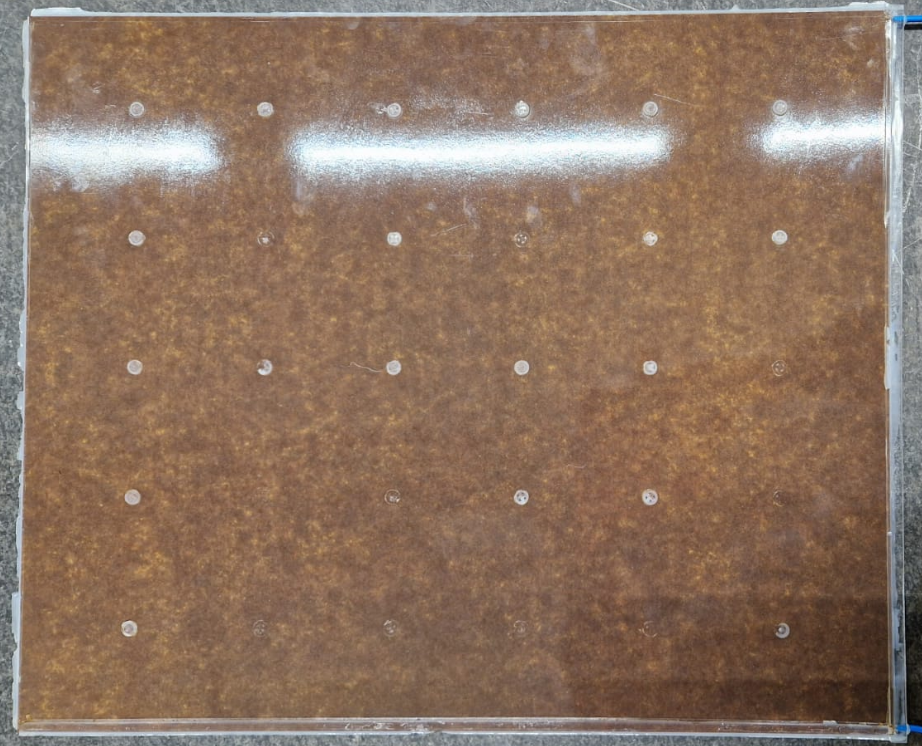}
  \caption{ }
  \label{fig:mockup_gas_volume_pic01}
\end{subfigure}
\hfill
\begin{subfigure}[b]{.49\textwidth}
  \includegraphics[width=\linewidth]{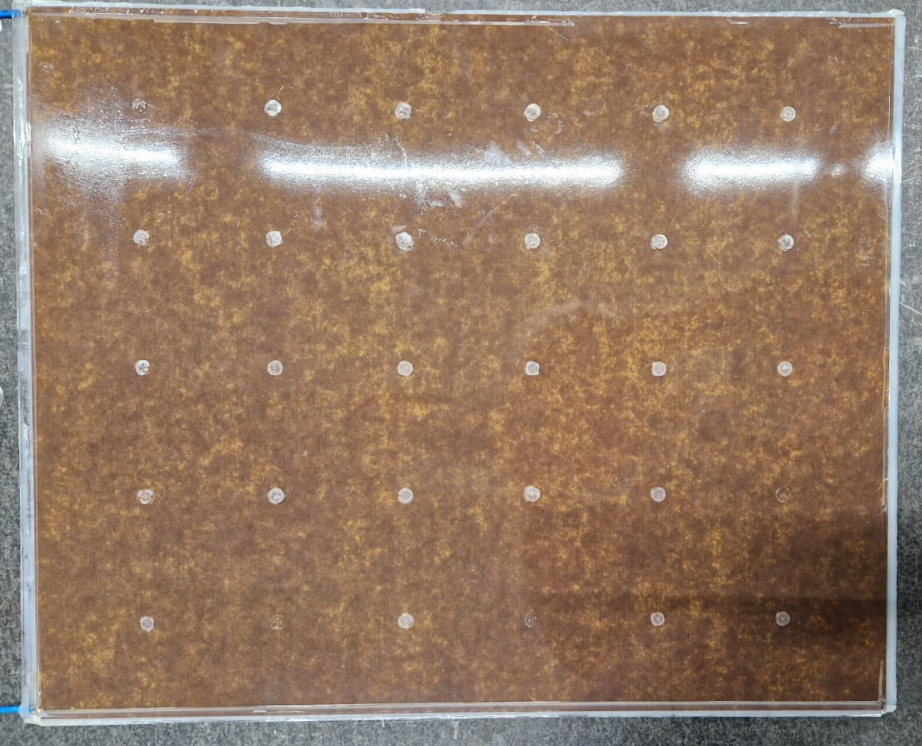}
  \caption{ }
  \label{fig:mockup_gas_volume_pic02}
\end{subfigure}

\vspace{6pt}

\begin{subfigure}[b]{.49\textwidth}
  \includegraphics[width=\linewidth]{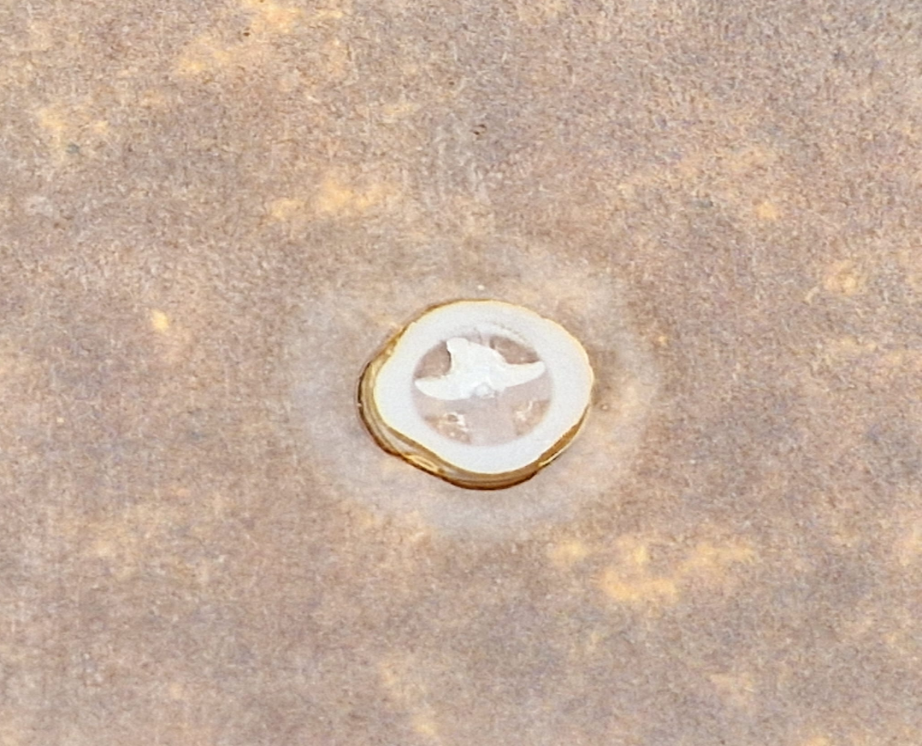}
  \caption{ }
  \label{fig:mockup_gas_volume_pic03}
\end{subfigure}
\hfill
\begin{subfigure}[b]{.49\textwidth}
  \includegraphics[width=\linewidth]{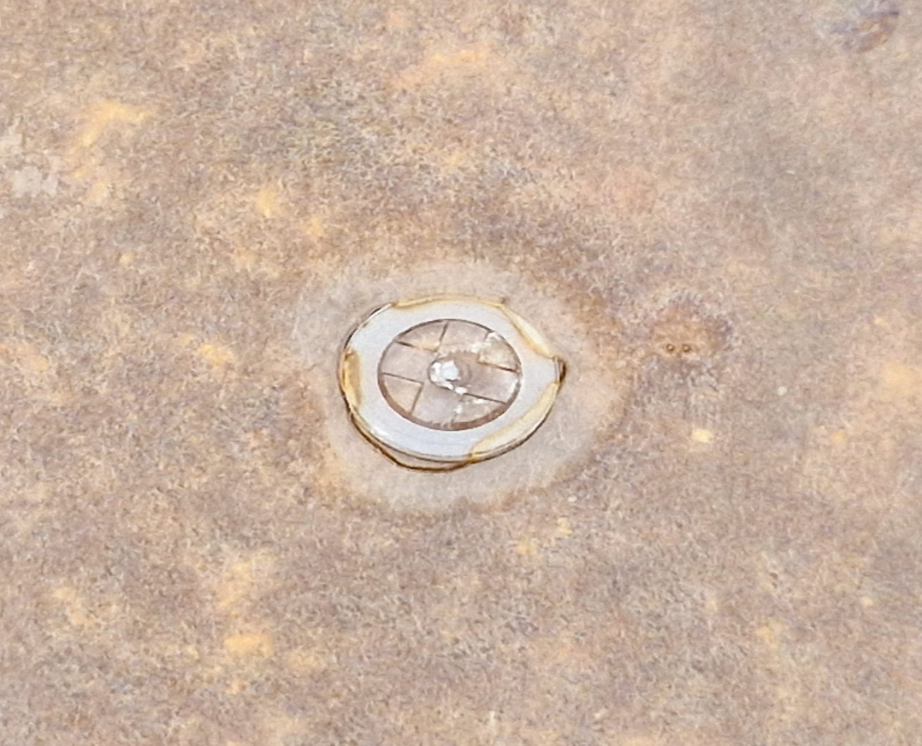}
  \caption{ }
  \label{fig:mockup_gas_volume_pic04}
\end{subfigure}
\caption{(a) and (b) Photographs document the opening of the mock-up gas volume and the ensuing visual inspection performed immediately after linseed-oil polymerization reached completion. (c) and (d) Enlarged detail views highlighting the halos surrounding the polycarbonate spacers.}
  \label{fig:2x2}
\end{figure}

A dedicated small-size mock‑up of the gas volume is manufactured to replicate the constituent materials, structural layout and, crucially, the linseed oil varnish coating applied to the inner surfaces of the HPL electrodes. This mock‑up gas volume undergoes the complete linseed oil protocol, application, draining, and controlled‑humidity curing, identical to that used for operational real-size gas volumes, thereby ensuring polymerization dynamics indistinguishable from those of the final gas volume. Once the curing cycle is complete, the mock-up gas volume is opened in a clean environment and the internal surface is examined through a dedicated quality‑assurance protocol: macroscopic visual inspection verified the uniformity and completeness of the cross‑linked oil film, while a quantitative scratch‑resistance test (10 mm blade, 1 N load, $45^\circ$ incidence, 100 mm track) probed its mechanical integrity. The absence of visible abrasions, delamination or particulate inclusions confirmed that the coating meets the stringent specifications required for stable long‑term detector operation. Figures \ref{fig:mockup_gas_volume_pic01} and \ref{fig:mockup_gas_volume_pic02} document the opening and subsequent visual inspection of a mock‑up gas volume carried out at the completion of linseed oil polymerization. The coating forms a continuous, uniformly distributed film on both HPL plates, with no discernible wet or tacky areas, confirming full oxidative cross‑linking. Localized halos, approximately $1.0-1.3 \, cm$ in diameter, are present around the spacers, as shown in Figures \ref{fig:mockup_gas_volume_pic03} and \ref{fig:mockup_gas_volume_pic04}; although these features do not compromise the detector performance, their formation mechanism warrants dedicated chemical follow‑up analyses.

\subsection{Volt-Amperometric characteristic test}
\label{subsec:Volt-AmperometricCharacteristicTest}
\begin{figure}[h]
    \centering
    \includegraphics[width=1\textwidth]{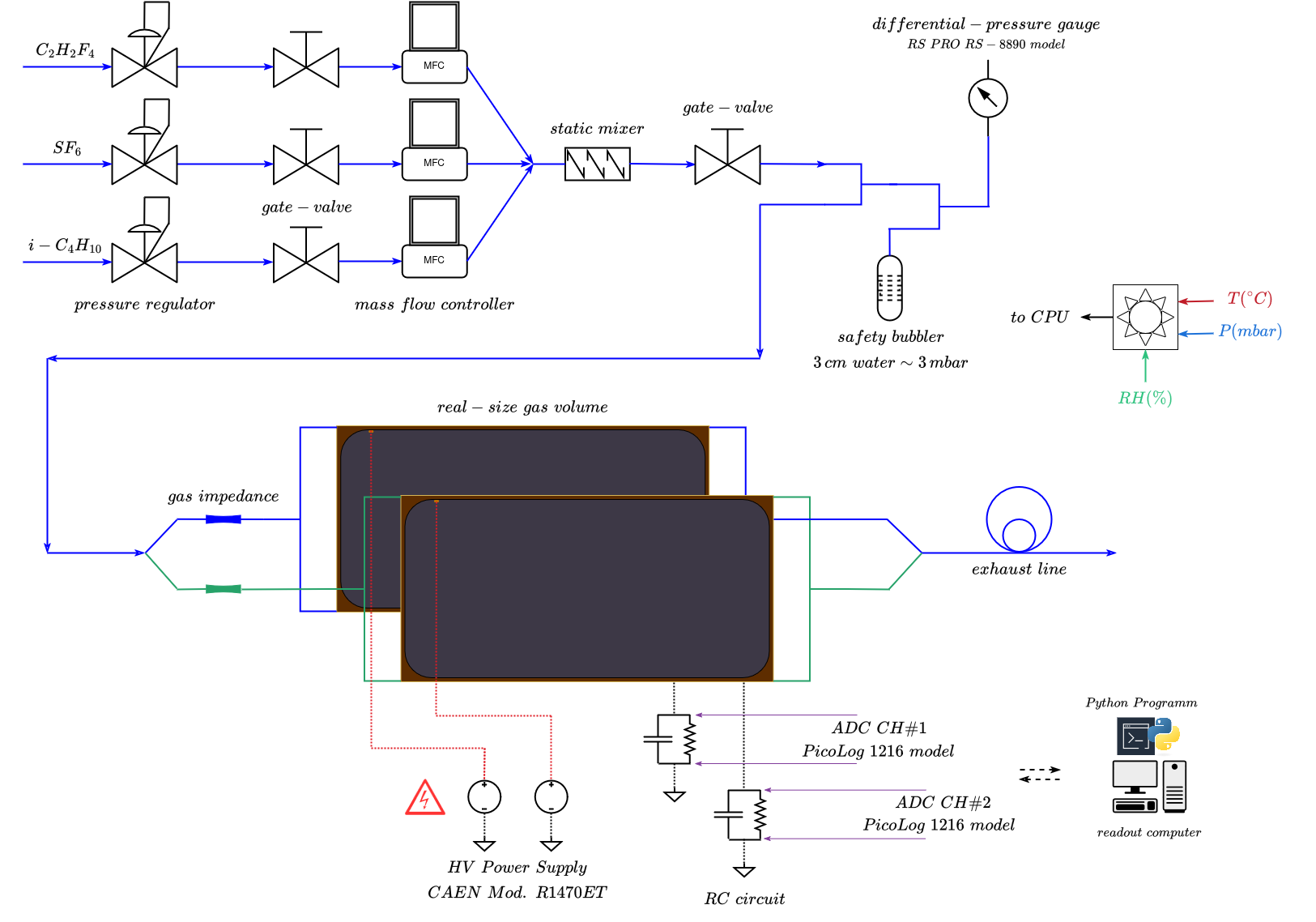} 
    \caption{Diagram of the test stand for the Volt-Amperometric characteristic test.}
    \label{fig:SchematicVolt-Amperometric}
\end{figure}

Each RPC gas volume is characterized by recording the Volt - Amperometric characteristic curve, which details the relationship between applied high voltage and the dark current (gap current). This test is conducted with the standard ATLAS-RPC gas mixture (94.7\% $C_2H_2F_4$ : 5\% $i-C_4H_{10}$ : 0.3\% $SF_6$). The gap current comprises two main components: an ohmic current that flows through the Bakelite electrodes, spacers, and frame, and an avalanche-induced current generated within the gas. The voltage scan is conducted by gradually increasing the applied voltage in 200 V increments up to 4 kV, followed by 100 V increments up to the maximum operating voltage of 6.2 kV. Current measurements are performed through a resistor connected between one of the HPL plates and ground, allowing accurate recording of the current flow via an Analog-to-Digital Converter (ADC). A diagram of the test stand used for Volt-Amperometric characteristic test of a full-scale RPC gas volume is shown in Figure \ref{fig:SchematicVolt-Amperometric}.

\begin{figure}[h]
    \centering
    \includegraphics[width=0.75\textwidth]{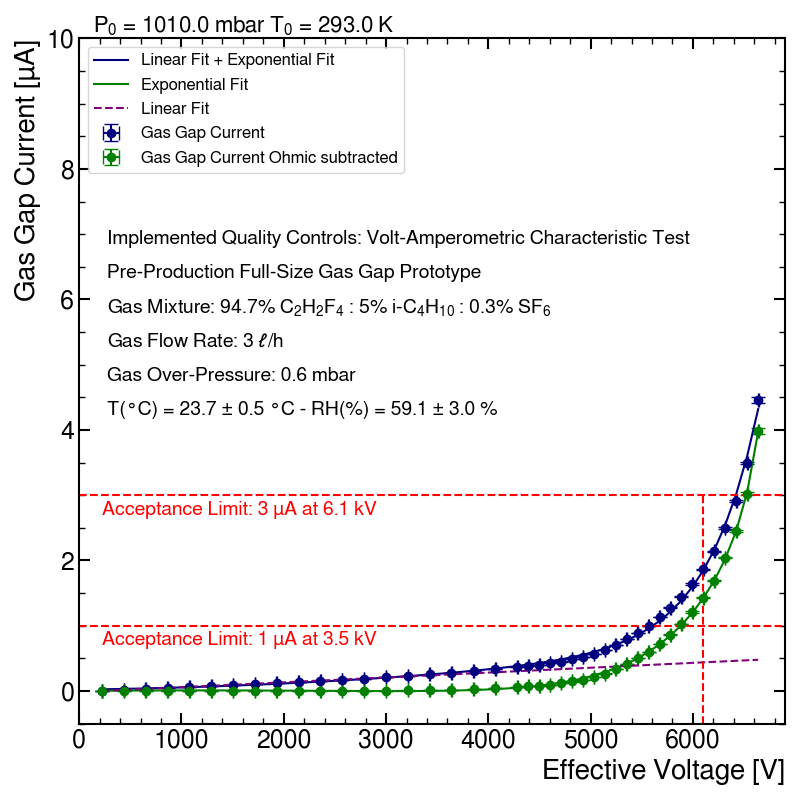} 
    \caption{Volt-Amperometric characteristic test for a full-scale RPC gas volume}
    \label{fig:Volt-AmperometricCharacteristicTest}
\end{figure}

The applied high voltage $V_{app}$ is corrected for local changes in environmental pressure P and temperature T at the production and certification facility by a high voltage correction factor to calculate the effective voltage $V_{eff}$ using the equation: 
\begin{equation}
V_{eff} = V_{app} \times \frac{P_0}{P} \times \frac{T}{T_0}
\label{eq:PTCorrection}
\end{equation}

\noindent where $P_0$ = 1010 mbar and $T_0$ = 293 K are the reference temperature and pressure. The gap current is modeled as:
\begin{equation}
I_{gap} = \frac{V_{eff}}{R_{bulk}} + I_0e^{\frac{V_{eff}}{V_0}}
\label{eq:PTCorrection}
\end{equation}

\noindent where $R_{bulk}$ represents the bulk ohmic resistance, $I_0$ is the mean primary charge produced per unit time, and $V_0$ depends on the gas mixture’s effective Townsend coefficient. In the range of 0–3.5 kV, the current trend is predominantly ohmic, reflecting the bulk resistance. Above 4 kV, the current’s exponential increase is driven by gas ionization effects. To meet quality standards, gas volumes are rejected if their leakage current exceeds 1 $\mu$A at 3.5 kV or if the exponential current exceeds 3 $\mu$A at 6.1 kV. This stringent criterion ensures that only volumes with minimal dark current and stable gas amplification advance to deployment. Representative results from the Volt-Amperometric characteristic test  are presented in Figure \ref{fig:Volt-AmperometricCharacteristicTest}.

\subsection{Long-term stability test}
The long-term stability test is an essential procedure for assessing and certifying the quality and reliability of RPC gas volumes. This test involves the continuous monitoring of the current drawn by the gas volume while it is subjected to a sequence of predefined high-voltage levels, using the standard gas mixture employed by the ATLAS RPC system. The high voltage is automatically ramped up in three sequential stages: initially from 0 kV to 5.0 kV, followed by an increase to 5.4 kV, and finally reaching the nominal operating voltage of 5.8 kV. The total duration of the test is two weeks. The intermediate voltage levels of 5.0 kV and 5.4 kV are each maintained for 48 hours to allow the system to stabilize. Subsequently, the high-voltage is set to 5.8 kV and held constant for the remainder of the testing period, thereby enabling a thorough investigation of the long-term behavior of the leakage current under sustained high-voltage stress. Throughout the entire test campaign, both the applied voltage and environmental parameters, including atmospheric pressure, ambient temperature, and relative humidity, are continuously monitored and logged. To mitigate the impact of environmental variations on the electrical response of the gas volume, the applied high-voltage is dynamically corrected in real time using the following standard formula \cite{Ballabene:2024qsz}:

\begin{equation}
\rho(P, T) = \left[ 1 + \alpha_P \left( \frac{P}{P_0} - 1 \right) \right] 
             \left[ 1 + \alpha_T \left( \frac{T_0 - 273.15}{T - 273.15} - 1 \right) \right].
\label{eq:PTCorrection2ndOrder}
\end{equation}

\noindent In the formula, $\alpha_P = 0.8$, $\alpha_T = 0.5$, $P_0 = 9.6 \times 10^4 \, Pa$, $T_0 = 294.16 \, K$ and $V_{ref} = 5.0, \, 5.4, \, 5.8 \, V$. The current drawn by the gas volume is precisely determined by measuring the voltage drop across a precision resistor connected between one of the HPL electrodes and ground. This configuration allows for accurate and stable current readout via an Analog-to-Digital Converter (ADC), as illustrated in Figure \ref{fig:SchematicVolt-Amperometric}. Given the extended duration of the stability test, a dedicated data acquisition system has been developed. This system ensures real-time data visualization, operational control, and robust logging of all relevant parameters. In the presence of anomalous conditions potentially compromising gas volume safety, the control system autonomously activates predefined interlock procedures to transition the gas volume into a safe operational state; for example, by disabling high-voltage channels in response to sudden current surges or reductions in gas flow rate.

\begin{figure}[h]
    \centering
    \includegraphics[width=1.0\textwidth]{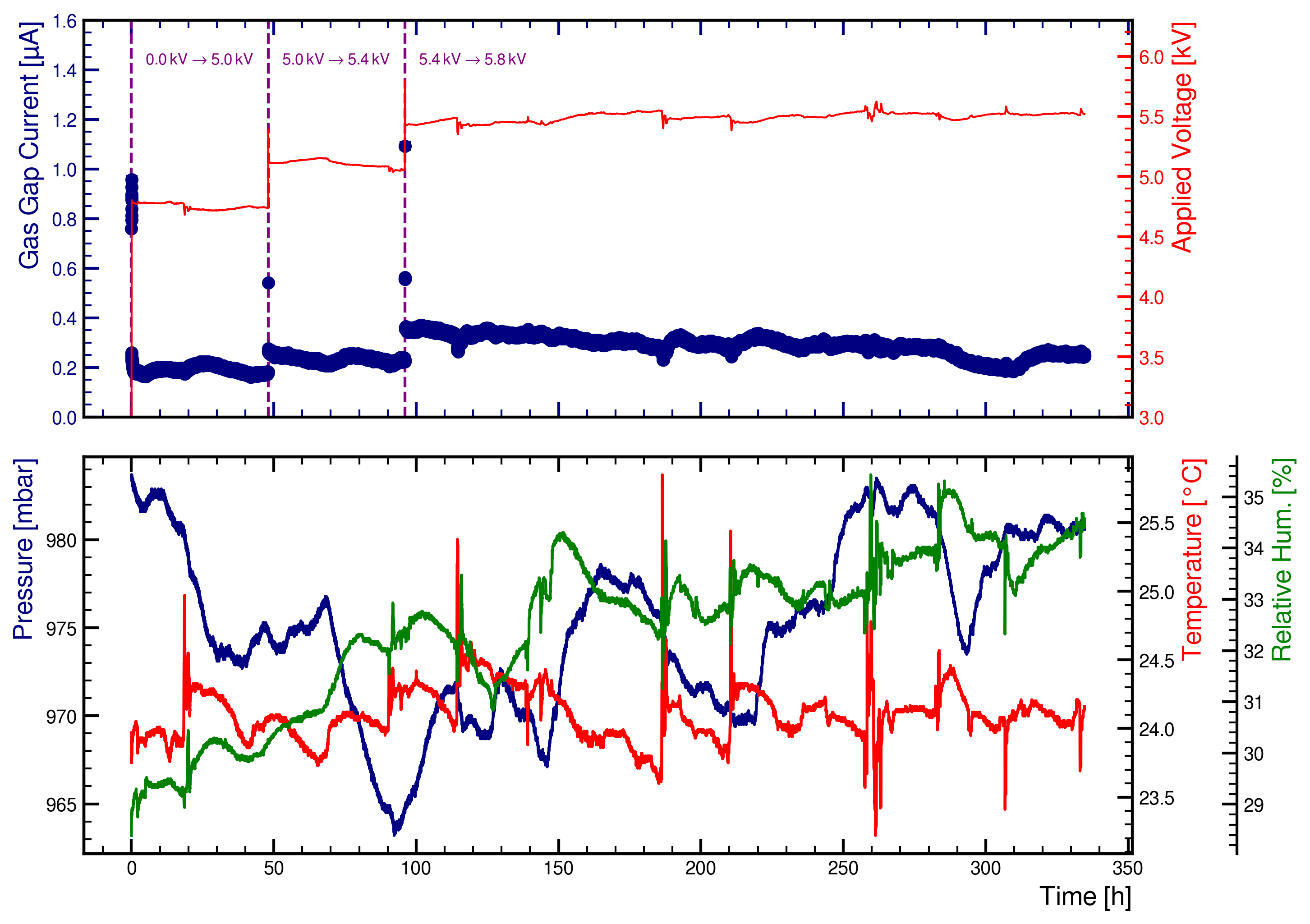} 
    \caption{Long-term stability test for a full-scale RPC gas volume}
    \label{fig:LongTermStabilityTest}
\end{figure}

Figure \ref{fig:LongTermStabilityTest} (top) presents an example of the current drawn by the gas volume as a function of time. Three distinct voltage windows are identifiable: the first lasting 48 hours at 5.0 kV, the second 48 hours at 5.4 kV, and the final one extending over 240 hours at 5.8 kV. A slow and continuous decrease in the current over time is also observed, attributed to the conditioning effect. Through this process, the resistive electrodes gradually reach a steady-state potential, while micro-impurities present within the gas volume are progressively eliminated through localized discharge processes. Figure \ref{fig:LongTermStabilityTest} (bottom) shows the environmental parameters, including atmospheric pressure, ambient temperature, and relative humidity, continuously monitored during the stability test. These values is employed to perform high-voltage corrections, ensuring consistent and reliable gas volume operation under varying environmental conditions. Compensating for environmental variations is critical to prevent them from masking potential deviations or early indicators of malfunction within the gas volume.

\begin{figure}[h]
    \centering
    \includegraphics[width=0.75\textwidth]{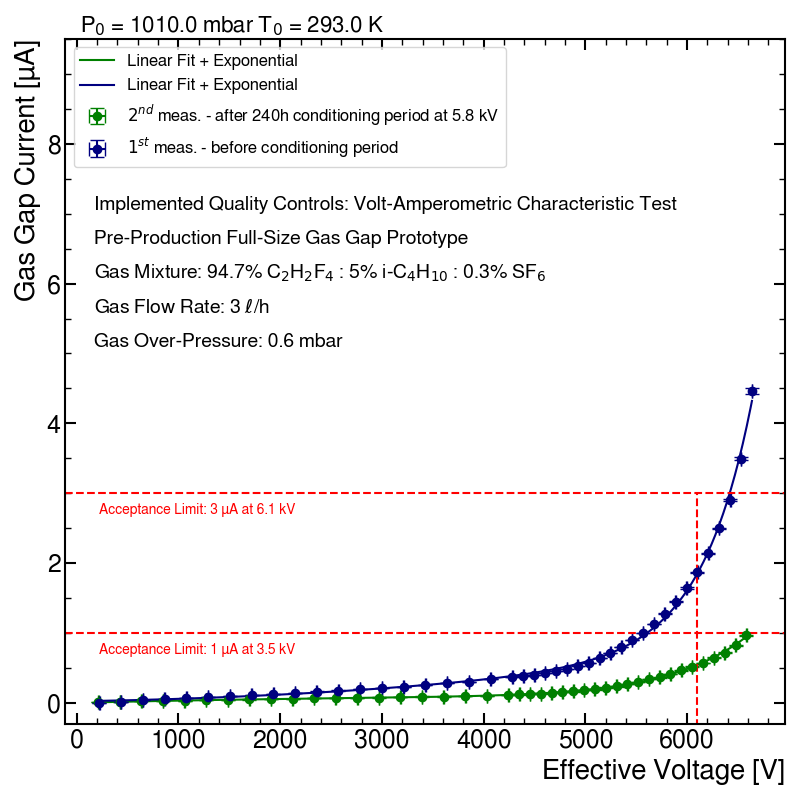} 
    \caption{Volt-Amperometric characteristic comparison for a full-scale gas volume, measured prior to and following the stability assessment and conditioning procedure.}
    \label{fig:Volt-AmperometricCharacteristicTestComparison}
\end{figure}

At the end of the stability test, a new measurement of the Volt-Amperometric characteristic curve is performed, following the procedure described in Section \ref{subsec:Volt-AmperometricCharacteristicTest}, and compared to the reference curve acquired prior to the stability test. Figure \ref{fig:Volt-AmperometricCharacteristicTestComparison} shows the comparison of the pre- and post-conditioning Volt-Amperometric characteristic curves. An overall improvement in the detector performance is observed, as evidenced by a systematic reduction in the measured currents across both the ohmic and avalanche regions. This behavior is consistent with the expected conditioning process, during which micro-impurities are gradually removed and the resistive electrodes approach a stable operating regime, resulting in reduced leakage and avalanche currents.

\subsection{Principal quality‑control test summary for full‑scale RPC gas volumes}
\begin{figure}[h]
    \centering
    \includegraphics[width=0.75\textwidth]{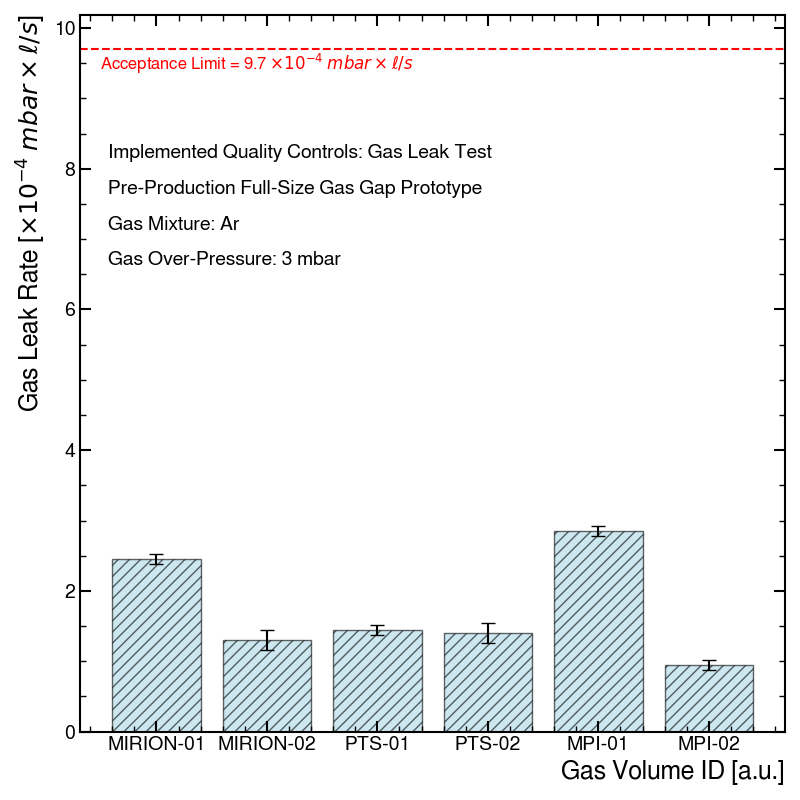} 
    \caption{Summary of results from argon-based gas-tightness measurements conducted on full-scale RPC gas volumes manufactured by MPI and its German industrial partners.}
    \label{fig:SummaryGasTightness}
\end{figure}

\begin{figure}[h]
    \centering
    \includegraphics[width=0.75\textwidth]{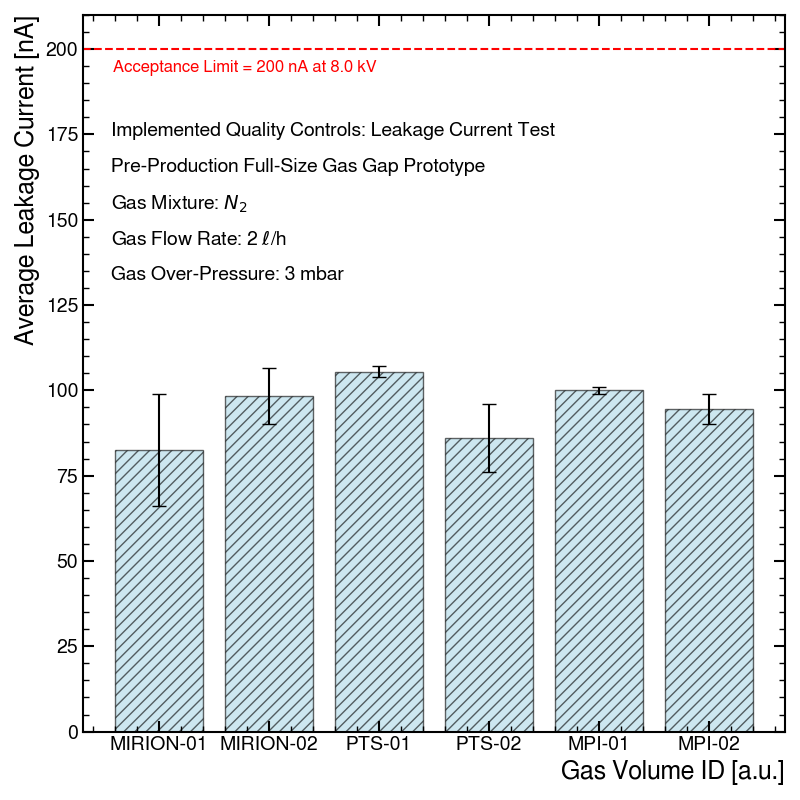} 
    \caption{Summary of leakage-current measurements performed on full-scale RPC gas volumes manufactured by MPI and its German industrial partners.}
    \label{fig:SummaryLeakageCurrent}
\end{figure}

\begin{figure}[h]
    \centering
    \includegraphics[width=0.75\textwidth]{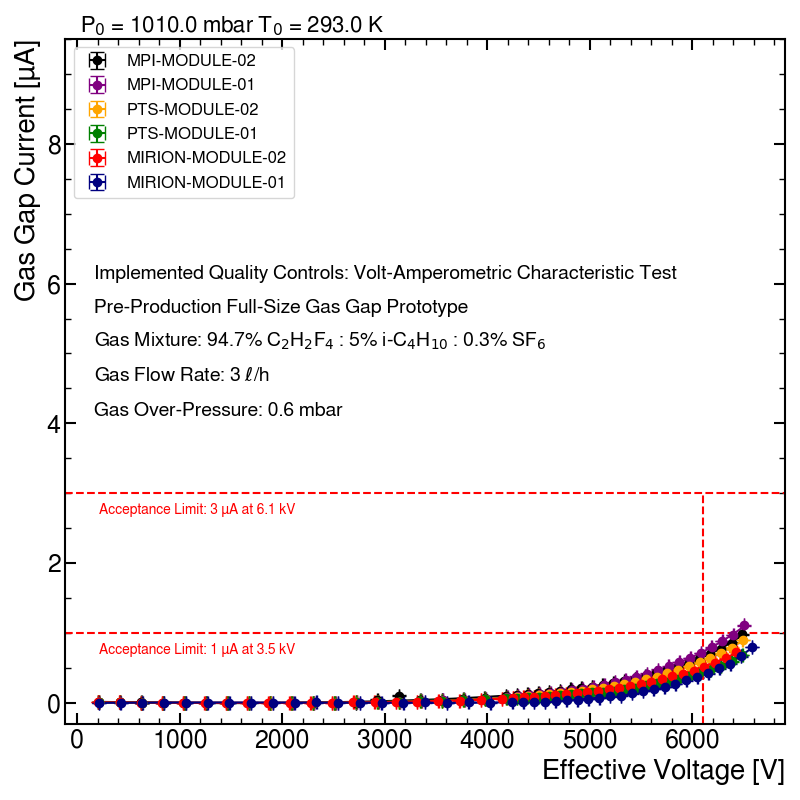} 
    \caption{Summary of Volt–Amperometric characterization of full-scale RPC gas volumes manufactured by MPI and its German industrial partners.}
    \label{fig:SummaryVoltAperometric}
\end{figure}

Comprehensive quality-control campaigns demonstrate that the full-scale RPC gas volumes produced at Max Planck Institute, in collaboration with German industrial partners, fully comply with the stringent performance requirements for application in high-energy physics experiments. Argon‑based gas‑tightness measurements give leak rates below $9.7 \times 10^{-4} \, mbar \times \ell / s$, thereby guaranteeing long‑term integrity over the detector operational lifetime. Complementary leakage‑current tests yield dark‑current below $200 \, nA$ at the maximum applied voltage of 8 kV on both HPL electrode, corresponding to a factor 1.4 above the nominal operating voltage of 5.8 kV, thereby confirming excellent surface insulation and the absence of conductive defects on the gas volume walls. Finally, the Volt–Amperometric characteristic tests of the RPC gas volumes exhibit an initial low-voltage ohmic regime, in which the measured current scales with the electrode resistivity, followed by a clear “knee” at the avalanche onset voltage. Beyond this threshold, the current undergoes a rapid exponential rise driven by gas-phase micro-discharges. The reproducibility of the knee position and the persistently low dark-current levels confirm that every gas volumes produced has met the stringent leak-tightness and operational-functionality criteria. Six full-scale RPC gas volumes have been successfully assembled, two at MIRION\textsuperscript{\tiny \textregistered} , two at PTS\textsuperscript{\tiny \textregistered} , and two at MPI, and all units manufactured by MPI and its German industrial partners meet the stringent acceptance criteria. Figures \ref{fig:SummaryGasTightness}, \ref{fig:SummaryLeakageCurrent} and \ref{fig:SummaryVoltAperometric} summarize the key quality-control assessments performed on these volumes: argon-based gas-tightness measurements, precision leakage-current scans, and detailed Volt–Amperometric characterizations. The full-scale gas volumes have been subsequently delivered to CERN and installed in the GIF++ irradiation facility for an accelerated radiation campaign, which is designed to validate both the integrity of the detector components and the robustness of the production procedures under operational conditions.

\section{RPC gas volume production facility at MPI}
\label{sec:MPIFacility}
\begin{figure}[htbp]
  \centering
  \begin{minipage}[c]{0.5\linewidth}
    \centering
    \begin{subfigure}[b]{\linewidth}
      \includegraphics[width=\linewidth]{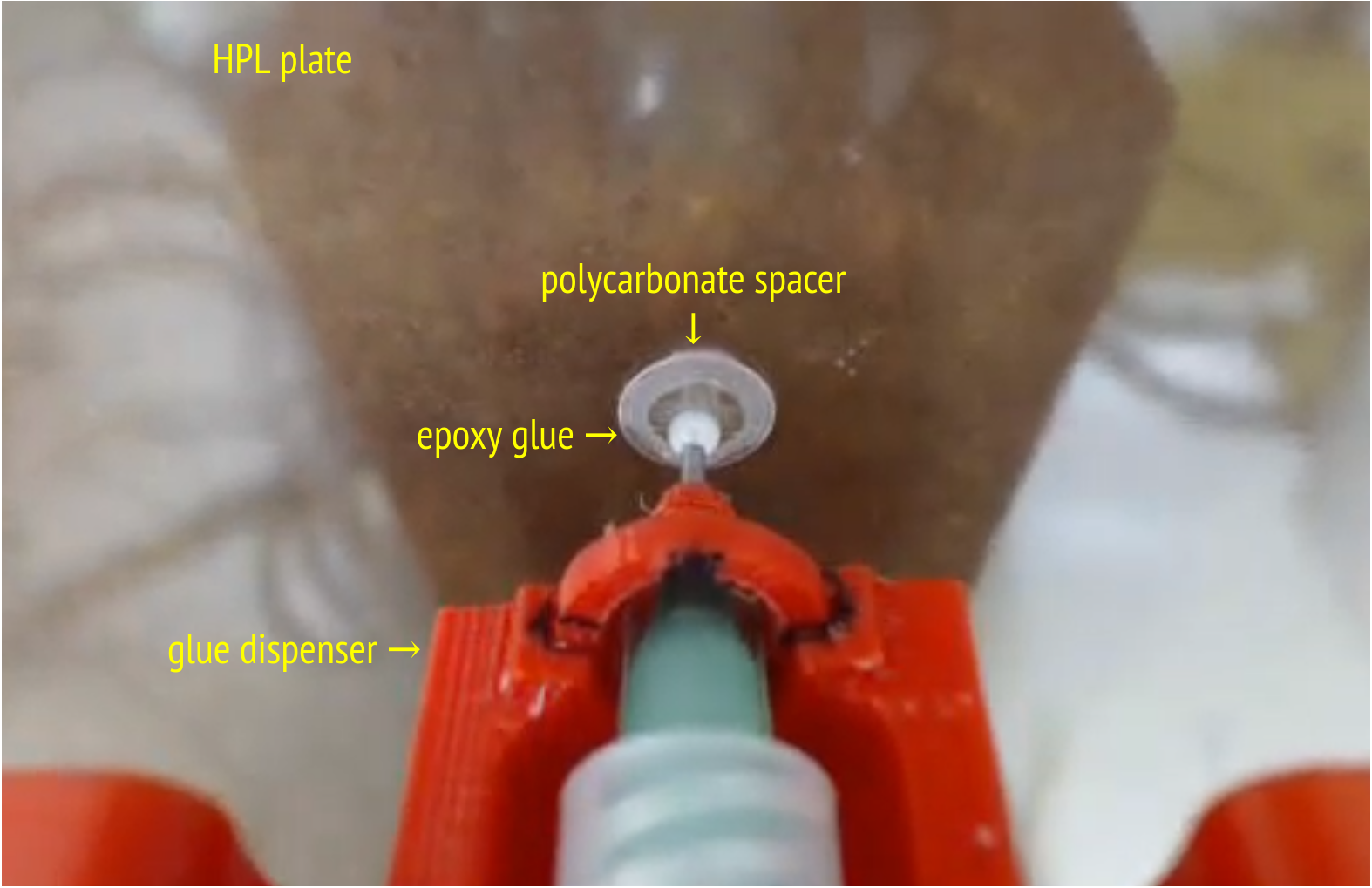}
      \caption{}    
      \label{fig:MPI_Gluing_Freeze_Frames_A}
    \end{subfigure}

    \vspace{0.6em}  

    \begin{subfigure}[b]{\linewidth}
      \includegraphics[width=\linewidth]{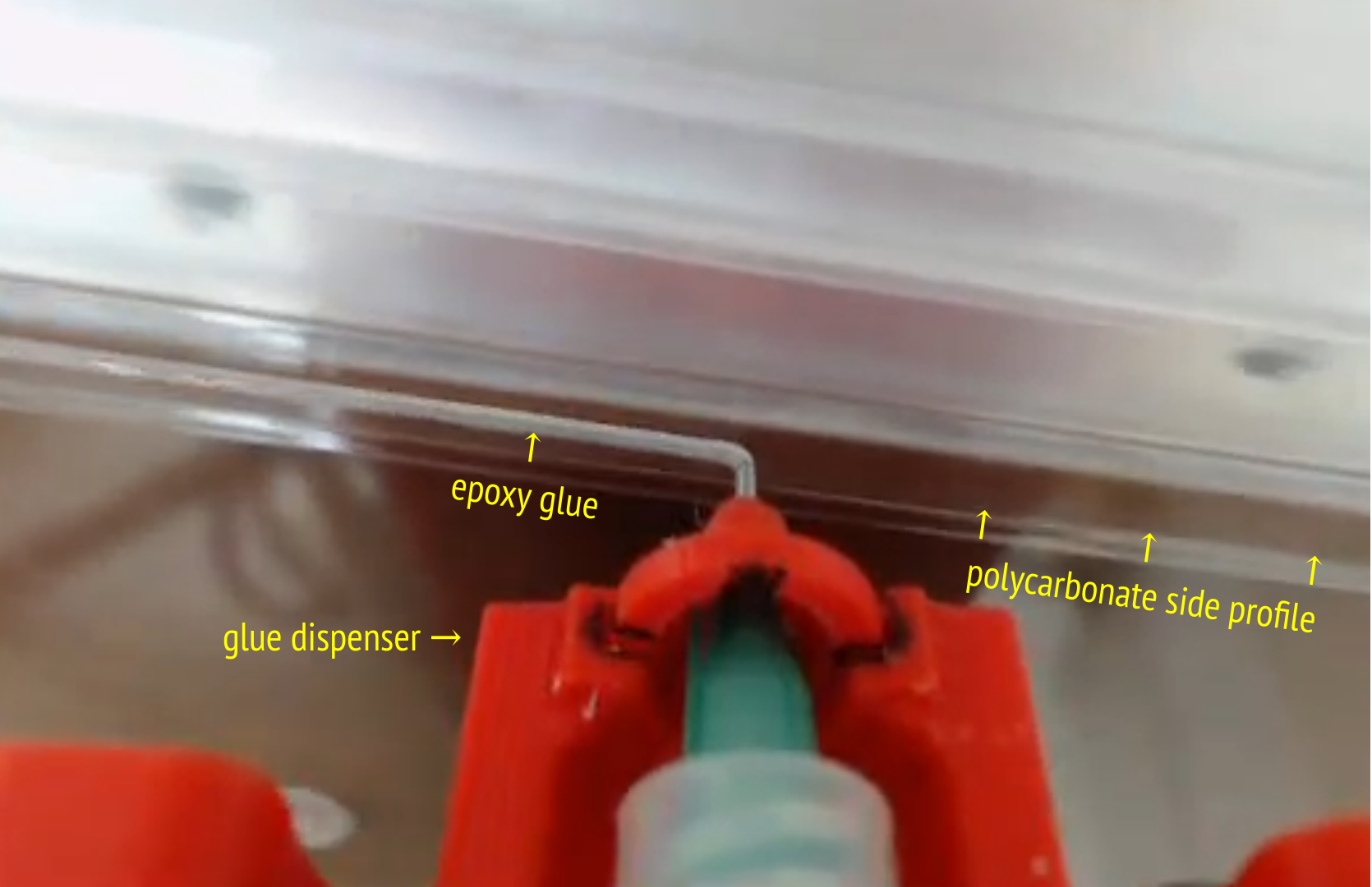}
      \caption{}    
      \label{fig:MPI_Gluing_Freeze_Frames_B}
    \end{subfigure}
  \end{minipage}
  \hfill
  \begin{minipage}[c]{0.45\linewidth}
    \centering
    \begin{subfigure}[b]{\linewidth}
      \includegraphics[width=\linewidth]{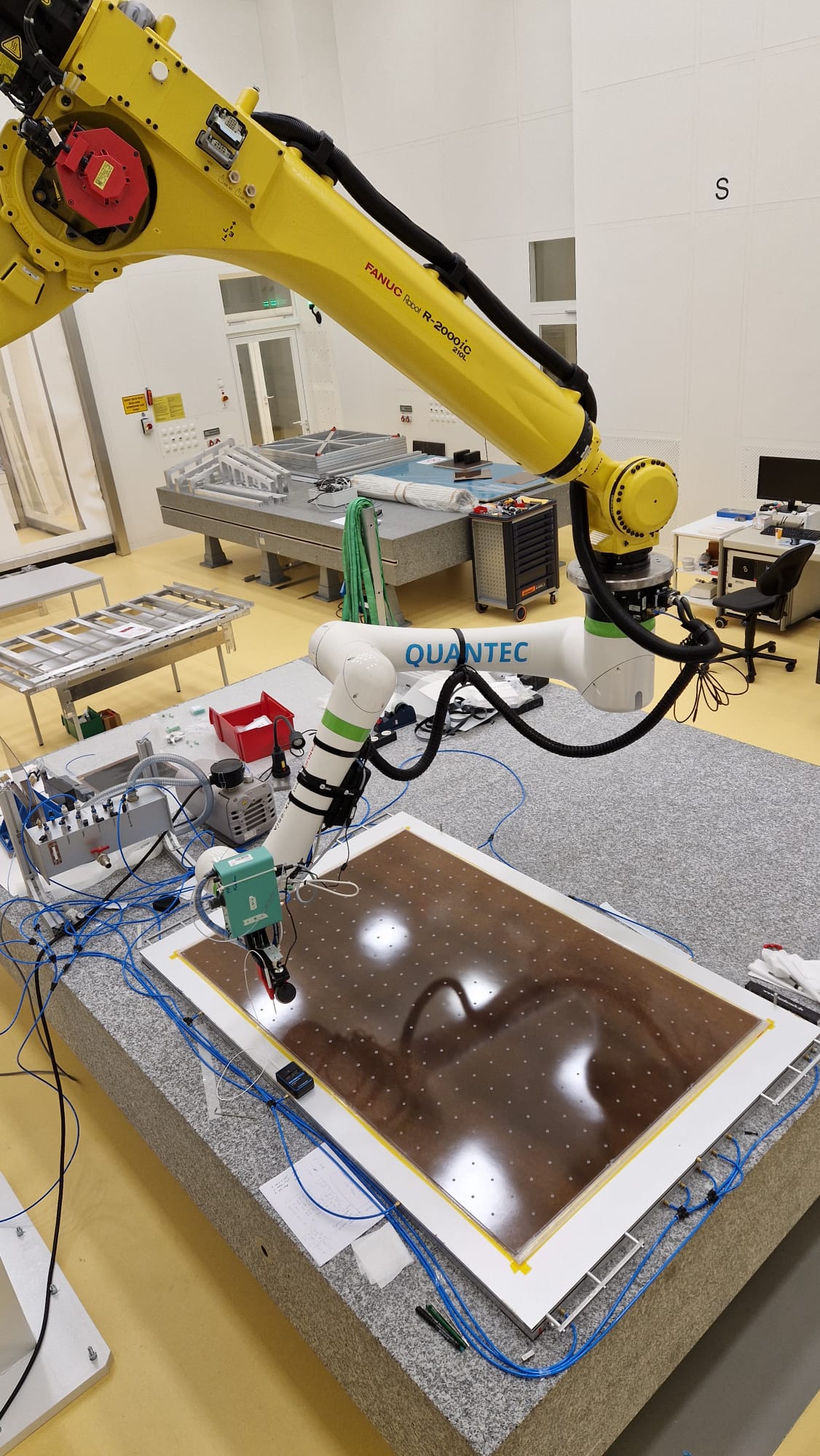}
      \caption{}    
      \label{fig:MPIFacility}
    \end{subfigure}
  \end{minipage}

  \caption{(a) and (b) Freeze frames capturing the glue dispensing onto the spacers (top left) and along the side profiles (bottom left) of the gas volume. (c) Assembly of a gas volume in the clean room at the Max Planck Institute. The controlled environment ensures precision and cleanliness in relevant production steps, critical to achieving high quality standards in gas volume construction.}
  \label{fig:three_photos}
\end{figure}

The production and certification facility for gas volumes has reached full operational status within the newly established clean room at the Max Planck Institute (MPI) building at the Garching Research Center. This state-of-the-art facility is designed to serve as a contingency site, ready to support or supplement production requirements should the primary German industrial facilities face any interruptions or increased demand. The MPI production center incorporates an advanced collaborative robotic system, enabling automation of critical steps in the gas volume assembly process. The robotic arm is equipped with integrated cameras, allowing real-time monitoring of the gluing phases directly at the application point. This video system detects any glue dispensing errors immediately, enabling prompt corrective actions to ensure high precision and reliability in assembly. Figures \ref{fig:MPI_Gluing_Freeze_Frames_A} and \ref{fig:MPI_Gluing_Freeze_Frames_B} show the freeze frames capturing the glue dispensing process during the assembly phase. The clean room provides a meticulously controlled environment, ensuring precise regulation of temperature and humidity - critical parameters for achieving high-quality gas volume bonding. Furthermore, this environment is indispensable for upholding the stringent standards of precision, cleanliness, and consistency required in gas volume assembly processes. Building on the expertise developed at MPI, industrial partners have been tasked with establishing an environment with controlled temperature and humidity to meet the stringent requirements of future series production of the gas volumes. Figure \ref{fig:MPIFacility} shows a photograph taken inside the clean room at the Max Planck Institute during the assembly phase of a gas volume.

\end{document}